\documentclass[oldversion,longauth]{aa}

\usepackage{subfigure}
\usepackage[dvips]{graphicx}
\usepackage[latin1]{inputenc}
\usepackage{amsmath}
\usepackage{amssymb}
\usepackage[round]{natbib}
\usepackage{verbatim}

\begin{document}

\title{Spot the difference}
\subtitle{Impact of different selection criteria on observed properties\\ 
of passive galaxies in zCOSMOS-20k\thanks{based on data obtained with the European Southern Observatory
Very Large Telescope, Paranal, Chile, program 175.A-0839} sample
}

\author{
M.~Moresco \inst{1} 
\and L.~Pozzetti \inst{2}
\and A.~Cimatti \inst{1} 
\and G.~Zamorani \inst{2}
\and M.~Bolzonella \inst{2}
\and F.~Lamareille \inst{3}
\and M.~Mignoli \inst{2}
\and E.~Zucca \inst{2}
\and S.J.~Lilly \inst{4}
\and C.M.~Carollo \inst{4}
\and T.~Contini \inst{3,5}
\and J.-P.~Kneib \inst{6}
\and O.~Le~F\`evre \inst{6}
\and V.~Mainieri \inst{7}
\and A.~Renzini \inst{8}
\and M.~Scodeggio \inst{9}
\and S.~Bardelli \inst{2}
\and A.~Bongiorno \inst{10}
\and K.~Caputi \inst{23}
\and O.~Cucciati \inst{2}
\and S.~de~la~Torre \inst{11}
\and L.~de~Ravel \inst{11}
\and P.~Franzetti \inst{9}
\and B.~Garilli \inst{9}
\and A.~Iovino \inst{12}
\and P.~Kampczyk \inst{4}
\and C.~Knobel \inst{4}
\and K.~Kova\v{c} \inst{4}
\and J.-F.~Le~Borgne \inst{3,5}
\and V.~Le~Brun \inst{6}
\and C.~Maier \inst{4,24}
\and R.~Pell\'o \inst{3,5}
\and Y.~Peng \inst{13}
\and E.~Perez-Montero \inst{3,5}
\and V.~Presotto \inst{12}
\and J.D.~Silverman \inst{14}
\and M.~Tanaka \inst{14}
\and L.~Tasca \inst{6}
\and L.~Tresse \inst{6}
\and D.~Vergani \inst{15}
\and L.~Barnes \inst{4}
\and R.~Bordoloi \inst{4}
\and A.~Cappi \inst{2}
\and C.~Diener \inst{4}
\and A.M.~Koekemoer \inst{13}
\and E.~Le Floc'h \inst{16}
\and C.~L\'opez-Sanjuan \inst{6,22}
\and H.J.~McCracken \inst{17}
\and P.~Nair \inst{2,13}
\and P.~Oesch \inst{4,18}
\and C.~Scarlata \inst{19}
\and N.~Scoville \inst{20}
\and N.~Welikala \inst{21}
}

   \offprints{Michele Moresco (\email{michele.moresco@unibo.it}) }

\institute{
Dipartimento di Fisica e Astronomia, Universit\'a degli Studi di Bologna, V.le Berti Pichat, 6/2 - 40127, Bologna, Italy
\and 
INAF -- Osservatorio Astronomico di Bologna, via Ranzani 1, I-40127, Bologna, Italy
\and 
Institut de Recherche en Astrophysique et Plan\'etologie, CNRS, F-31400 Toulouse, France
\and
Institute for Astronomy, ETH Zurich, 8093 Zurich, Switzerland
\and 
IRAP, Universit\'e de Toulouse, UPS-OMP, Toulouse, France
\and
Laboratoire d'Astrophysique de Marseille, Aix Marseille Universit\'e, CNRS, 13388, Marseille, France
\and 
European Southern Observatory, Garching, Germany
\and
INAF -- Osservatorio Astronomico di Padova, Padova, Italy
\and 
INAF -- Istituto di Astrofisica Spaziale e Fisica Cosmica, Milano, Italy
\and
INAF -- Osservatorio Astronomico di Roma, 00040, Monteporzio Catone, Italy
\and
Institute for Astronomy, The University of Edinburgh, Royal Observatory, Edinburgh, EH93HJ, UK
\and
INAF -- Osservatorio Astronomico di Brera, Milano, Italy
\and
Space Telescope Science Institute, Baltimore, MD 21218, USA
\and 
Kavli Institute for the Physics and Mathematics of the Universe (WPI), Todai Institutes for Advanced Study, The University of Tokyo, 5-1-5 Kashiwanoha, Kashiwa, 277-8583, Japan
\and 
INAF -- Istituto di Astrofisica Spaziale e Fisica Cosmica, Bologna, Italy
\and
Institute for Astronomy, University of Hawaii, 2680 Woodlawn Drive, Honolulu, HI 96822, USA
\and 
Institut d'Astrophysique de Paris, Universit\'e Pierre \& Marie Curie, 75014 Paris, France
\and 
UC Santa Cruz/UCO Lick Observatory, 1156 High Street, Santa Cruz, CA 95064, USA
\and
Minnesota Institute for Astrophysics, School of Physics and Astronomy, University of Minnesota, Minneapolis, MN 55455, USA
\and
California Institute of Technology, MC 249-17, 1200 East California Boulevard, Pasadena, CA 91125, USA
\and
Institut d'Astrophysique Spatiale, Batiment 121, CNRS \& Univ. Paris Sud XI, 91405 Orsay Cedex, France
\and
Centro de Estudios de F\'{\i}sica del Cosmos de Arago\'on, Plaza San Juan 1, planta 2, 44001, Teruel, Spain
\and
Kapteyn Astronomical Institute, University of Groningen, P.O. Box 800, 9700 AV Groningen, The Netherlands
\and
University of Vienna, Department of Astrophysics, Tuerkenschanzstrasse 17, 1180 Vienna, Austria
}

\authorrunning {M. Moresco, et al.}

\titlerunning {Dependence of passive galaxies properties on selection criterion}

\date{Received -- -- ----; accepted -- -- ----}

\abstract{
{\small
{\it Aims.} We present the analysis of photometric, spectroscopic, and morphological properties for differently selected 
samples of passive galaxies up to $z=1$ extracted from the zCOSMOS-20k spectroscopic survey. This analysis
intends to explore the dependence of galaxy properties on the selection criterion adopted, study the degree of
contamination due to star-forming outliers, and provide a comparison between different commonly used selection 
criteria. This work is a first step to fully investigating the selection effects of passive galaxies for future massive surveys such as Euclid.\\ 
{\it Methods.} We extracted from the zCOSMOS-20k catalog six different samples of passive galaxies, based on
morphology (3336 ``morphological'' early-type galaxies), optical colors (4889 ``red-sequence'' galaxies and 
4882 ``red UVJ'' galaxies), specific star-formation rate (2937 ``quiescent'' galaxies), a best fit to the observed 
spectral energy distribution (2603 ``red SED'' galaxies), and a criterion that combines 
morphological, spectroscopic, and photometric information (1530 ``red \& passive early-type galaxies''). For all the samples, 
we studied optical and infrared colors, morphological properties, specific star-formation rates, and the equivalent widths of
the residual emission lines; this analysis was performed as a function of redshift and stellar mass to inspect further
possible dependencies.\\ 
{\it Results.} We find that each passive galaxy sample displays a certain level of contamination due to blue/star-forming/nonpassive outliers. 
The morphological sample is the one that presents the higher percentage of contamination, with $\sim$12-65\% (depending on the mass range)
of galaxies not located in the red sequence, $\sim$25-80\% of galaxies with a specific star-formation rate up to $\sim$25 times
higher than the adopted definition of passive, and significant emission lines found in the median stacked spectra, at
least for ${\rm log({\mathcal M}/{\mathcal M}_{\odot})<10.25}$. The red \& passive ETGs sample is the purest, with a 
percentage of contamination in color $<$10\% for stellar masses ${\rm log({\mathcal M}/{\mathcal M}_{\odot})>10.25}$, 
very limited tails in sSFR, a median value $\sim$20\% higher than the chosen passive cut, and 
equivalent widths of emission lines mostly compatible with no star-formation activity. However, it is also the less economic 
criterion in terms of information used. Among the other criteria, we found that the best performing are the red
SED and the quiescent ones, providing a percentage of contamination only slightly higher than the red \& passive ETGs
criterion (on average of a factor of $\sim$2) but with absolute values of the properties of contaminants still compatible
with a red, passively evolving population. We also find a strong dependence of the contamination on the stellar mass
and conclude that, almost irrespective of the adopted selection criteria, a cut at ${\rm log({\mathcal M}/{\mathcal M}_{\odot})>10.75}$ 
provides a significantly purer sample in terms of star-forming contaminants. By studying the restframe color-mass and color-color
diagrams, we provided two revised definitions of passive galaxies based on these criteria that better reproduce the observed
bimodality in the properties of zCOSMOS-20k galaxies.\\
The analysis of the number densities of the various samples shows evidences of mass-assembly ``downsizing'', with galaxies
at ${\rm 10.25<log({\mathcal M}/{\mathcal M}_{\odot})<10.75}$ increasing their number by a factor $\sim$2-4 from $z=0.6$ to
$z=0.2$, by a factor $\sim$2-3 from $z=1$ to $z=0.2$ at ${\rm 10.75<log({\mathcal M}/{\mathcal M}_{\odot})<11}$, and by 
only $\sim$10-50\% from $z=1$ to $z=0.2$ at ${\rm 11<log({\mathcal M}/{\mathcal M}_{\odot})<11.5}$.

\keywords{galaxies: evolution -- galaxies: fundamental parameters -- galaxies: statistics -- surveys}
}
}

\maketitle

\section{Introduction}\label{sec:intro}
Early-type galaxies (ETGs) represent a population of galaxies apparently simple and homogeneous 
in terms of morphology, colors, stellar population content, and scaling relations \citep[for a detailed 
review, see][ and references therein]{Renzini2006}, at least in the local Universe.
Originally \citep{Hubble1936}, this population of galaxies was identified just on the basis of its morphology, with early-type galaxies
having a rounder, elliptical shape and late-type galaxies being characterised by spiral features.
Very soon, however, it was evident that this morphological dichotomy also reflected a difference in terms of the stellar population 
content and that the two classifications (morphological and based on the stellar population content) correlate, even if they were far from overlapping.

It has been evident for almost 50 years now that galaxy rest-frame colors show a clear bimodal distribution, both in clusters \citep[e.g.,][]{Visvanathan1977,Tully1982} 
and in the field \citep{Strateva2001,Hogg2002,Baldry2004a,Baldry2004b,Bell2004, Franzetti2007}.
Later studies have also shown a bimodality in many other galaxy parameters: $\rm{H\alpha}$ \citep{Balogh2004} and $\rm{[OII]}$
\citep{Mignoli2009} emission, 4000~\AA~ break \citep{Kauffmann2003}, star-formation history \citep{Brinchmann2004}, 
and clustering \citep{Meneux2006}. A similar bimodality has also been found in the galaxy stellar mass function, showing
that early-type galaxies are the most massive galaxies at $z\sim0$ \citep{Baldry2004a,Baldry2006,Baldry2008} and 
dominate the massive end of the stellar mass function up to $z\sim1$ \citep{Pozzetti2010}.

The properties of ETGs make them the ideal candidates to also probe cosmology, together with galaxy formation models and theories. 
These galaxies have been found to have a stronger clustering with respect to late-type galaxies 
\citep[e.g., see][and references therein]{Zehavi2011}, providing a better tracer to the structure of underlying matter.
It has also been proposed to use this population of galaxies as ``cosmic chronometers'', able to trace the differential 
age evolution of the Universe as a function of redshift \citep{Jimenez2002} and provide independent and precise measurements
of the Hubble parameter $H(z)$ up to $z\sim1.5$ \citep{Simon2005,Stern2010,Moresco2012a}. These measurements have been demonstrated 
to be an innovative and complementary tool to constrain cosmological parameters 
\citep{Moresco2011,Moresco2012b,Wang2012,Riemer2013,Aviles2013,Said2013}.

A powerful tool to study the evolution of a galaxy population is analyzing the evolution of its number density as a 
function of redshift, because this can give hints of the processes and characteristic timescales involved. The number
density of ETGs has been deeply studied by many authors, and there is a general agreement that, while
there is a strong evolution for less massive galaxies, the number density of ${\rm log({\mathcal M}/{\mathcal M}_{\odot})>11}$
ETGs is almost unchanged from $z\sim1$ to $z\sim0$ \citep{Cimatti2006,Scarlata2007b,Pozzetti2010,Brammer2011,Maraston2012,Ilbert2013,Moustakas2013}.
This has been interpreted as evidence that this population of galaxy has been already in place since $z\sim1$.

These pieces of observational evidence have induced a gradual shift of notation, so that with time the definition of ``early-type'' has been not only
related to a morphological feature but also to peculiar properties in terms of the stellar population content.
The ETGs are now usually identified as spheroidal (E/S0) galaxies 
with old stellar population, no (or negligible) star formation, and a passive evolution as a function of cosmic time.
Recently, all these properties are often used without a clear distinction, and ``red'', ``quiescent'', and ``old'' 
are erroneously used as synonyms. Surely, as discussed above, the bimodality in many galaxy properties highlights that 
red galaxies are mostly quiescent and passively evolving; however, a detailed and quantitative estimate of the contamination
due to blue/star-forming outliers should be made before doing a one-to-one correspondence. \cite{Franzetti2007} pointed 
out such possible contamination when studying a color-selected sample of ETGs, concluding  that selecting galaxies only 
on the basis of their colors can be misleading in estimating the evolution of old and passively evolving galaxies.

In this paper, we want to thoroughly explore what kind of galaxies are culled by different selection criteria, either purely 
morphological or based on the stellar population content as characterized by colors or star-formation rate (SFR). In particular, we shall 
quantify to which extent these criteria overlap in terms of selected galaxies and to which extent they differ.
Various global properties of the different samples are presented and discussed. In Sect. \ref{sec:data} we present our data
and how the different samples of passive galaxies have been selected. In Sect. \ref{sec:analysis} the color, spectroscopic, and
morphological analysis are presented. In Sect. \ref{sec:res} we quantify how much each sample is 
contaminated by the presence of blue/star-forming outliers and how this contamination depends on the stellar mass and 
on the adopted selection criterion. We also analyze the number densities of the various samples and compare
them to check for differences and provide insights about the characteristic processes involved. Finally, we
also present two revised color-mass and color-color selection criteria of passive galaxies, aiming to reduce the contamination by blue 
star-forming galaxies.

This study is interesting in the view of many outcoming surveys \citep[e.g., Euclid,][]{Laureijs2011} since it provides
a comparison between different methods of selecting passive galaxies, quantifying the purity of the sample as
well as the cost in terms of information needed.

Throughout the paper, we adopt the cosmological parameters $H_{0}=70$ ${\rm km}$ ${\rm s^{-1}}$ ${\rm Mpc^{-1}}$, $\Omega_{m}=0.25$, 
$\Omega_{\Lambda}=0.75$. Magnitudes are quoted in the AB system.

\section{Data}\label{sec:data}
\subsection{The sample}\label{sec:sample}
The COSMOS Survey \citep{Scoville2007} has imaged a field of $\sim 2$ deg$^2$ with the 
Advanced Camera for Surveys (ACS) with single-orbit I-band exposures to a depth $I_{AB}\simeq28$ mag 
and $50\%$ completeness at $I_{AB}=26.0$ mag for sources $0.5''$ in diameter \citep{Koekemoer2007}. 

The analysis presented in this paper is based on the zCOSMOS spectroscopic survey \citep{Lilly2007,Lilly2009}. 
This ESO Large Programme ($\sim 600$ hours of observations) was aiming to map the COSMOS 
field with the VIsible Multi-Object Spectrograph \citep[VIMOS,][]{LeFevre2003}, mounted on the ESO Very Large 
Telescope (VLT). 
Our sample has been extracted from the zCOSMOS 20k bright sample \citep{Lilly2009}. The observed magnitudes 
in 12 photometric bands (CFHT $u^*$, $K$ and $H$, Subaru $B_J$, $V_J$, $g^+$, $r^+$, $i^+$, and $z^+$, 
UKIRT $J$ and Spitzer IRAC at 3.6 $\mu$m and 4.5 $\mu$m) have been used in order to derive reliable estimates 
of galaxy parameters from the photometric SED fitting. The photometric catalog is described in \cite{Capak2007}. 
Following their approach, magnitudes were corrected for Galactic extinction using Table 11 of \cite{Capak2007}, 
and the photometry was optimized by applying zeropoint offsets to the observed magnitudes to reduce differences 
between the observed and reference magnitudes computed from a set of template spectral energy distributions (SEDs). 
The spectra have a medium resolution grism ($R\approx600$) with a slit width of 1 arcsec; the 
spectral ranges of the observations are typically $5550-9650$ {\AA}. The data were reduced with VIMOS Interactive 
Pipeline Graphical Interface software \citep[VIPGI,][]{Scodeggio2005} and the redshift measured with EZ software \citep{Garilli2010} 
with a high success rate \citep[$\approx95\%$ in the redshift range $0.5<z<0.8$,][]{Lilly2009}.
The line measurements were performed 
using the program {\it Platefit} \citep[see][for further details]{Lamareille2006}.
After removing the spectroscopically confirmed stars, the broad-line AGNs, and the galaxies with an insecure redshift measurement
($z_{\rm flag}<1.5$), we end up with a parent sample of 17211 galaxies.

The stellar masses (${\mathcal M}$) and SFRs were estimated for the entire sample by performing 
a best fit to the SEDs, using the observed magnitudes in 12 photometric bands from $u^*$ to $4.5\mu m$. A grid of 
theoretical models was built from \cite{BC03} stellar population synthesis models (hereafter BC03), adopting an 
exponentially declining star-formation history (SFH) with $\tau=0.1,0.3,0.6,1,2,3,5,10,15,30$ 
Gyr, a Chabrier initial mass function \citep{Chabrier2003}, a solar stellar metallicity, and a Calzetti extinction law \citep{Calzetti2000} 
with $0<A_{V}<3$. Absolute magnitudes and colors were evaluated using the ``algorithm for luminosity function'' (ALF) software following the method described in 
\cite{Zucca2009}. The specific Star Formation Rate (sSFR) has been defined as the ratio between the SFR
and the stellar mass, ${\rm sSFR=SFR/{\mathcal M}}$.

\begin{figure}[t!]
\begin{center}
\includegraphics[angle=-90, width=0.49\textwidth]{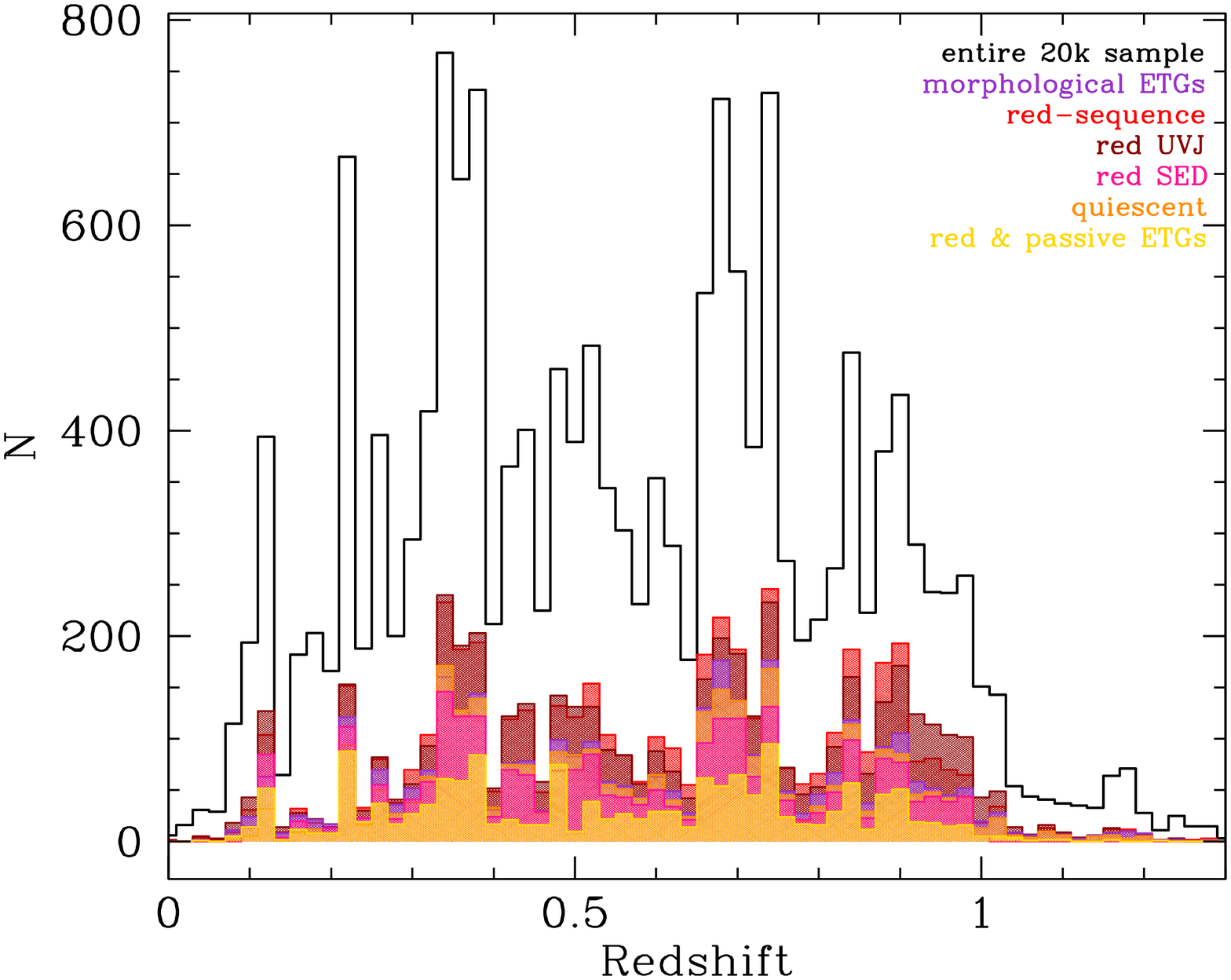}
\includegraphics[angle=-90, width=0.49\textwidth]{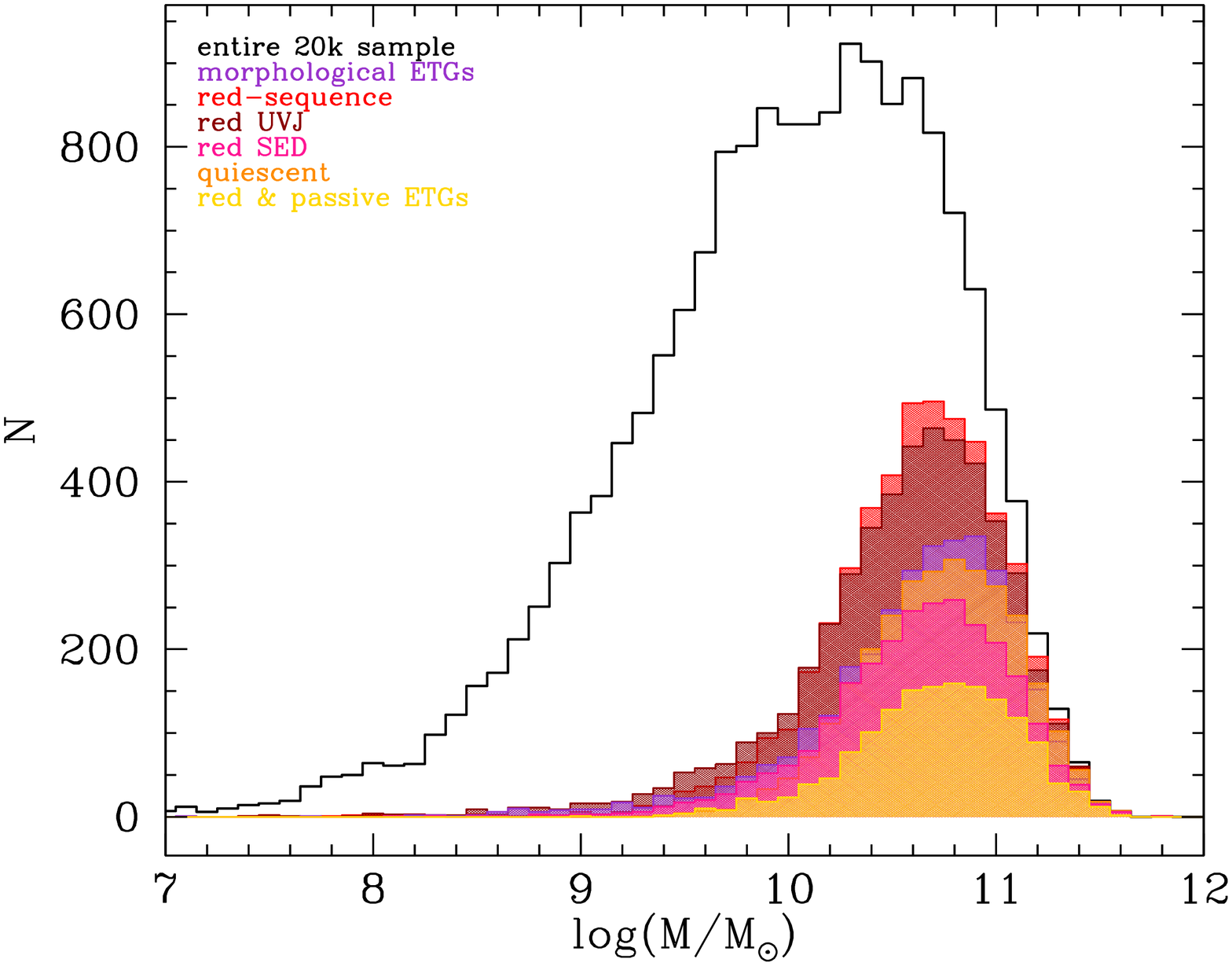}
\end{center}
\caption{Redshift distribution (upper panel) and stellar mass distribution (lower panel)
of the entire zCOSMOS-20k sample and of the different sample. In violet the morphological ETGs, 
in light red the red-sequence galaxies, in dark red the red UVJ galaxies, in pink the red SED 
galaxies, in orange the quiescent galaxies, and in yellow the red \& passive ETGs. 
\label{fig:ETGshist}}
\end{figure}

\subsection{The selection criteria}\label{sec:criteria}
Different selection criteria have been proposed so far to select passive galaxies.
We exploit here six different definitions, which are described below. 
These criteria were chosen so as not to be too restrictive because otherwise we would have 
poor completeness, but at the same time to minimize the contamination by blue, 
star-forming objects.

\begin{itemize}
\item {\it \bf ``morphological'' ETGs.} The Z\"urich Estimator of Structural Types \citep[ZEST;][]{Scarlata2007a}
estimates the morphological type of galaxies with a combination of a principal component analysis (PCA) 
of five nonparametric diagnostics of galaxy structure (asymmetry, concentration, Gini coefficient, second-order 
moment of the brightest 20\% galaxy pixels, and ellipticity) and of a parametric description of the galaxy light
(the exponent of a single-S\'ersic fit to the surface brightness distribution). This estimator was applied
to the full zCOSMOS-20k sample, providing morphological measurements for all the galaxies. We refer to \cite{Scarlata2007a}
for a discussion concerning the classification uncertainties using this method. Following this
classification, we selected 3336 ETGs matching an E/S0 morphology \citep{Scarlata2007a}.
\item {\it \bf ``red-sequence'' galaxies.} In their analysis, \cite{Peng2010} defined a color-mass relation calibrated on Sloan Digital Sky Survey (SDSS) and zCOSMOS-10k data, 
dividing red-sequence from blue-cloud galaxies. This relation also takes the redshift evolution into account and is a weak 
function of mass. Following their approach, 4889 red galaxies were selected on the basis of the following dividing 
rest-frame $(U-B)$ color.  
$$(U-B)_{\rm rest}>1.10+0.075\times log(\frac{\mathcal M}{10^{10}\mathcal M_{\odot}})-0.18\times z$$
\item {\it \bf ``red UVJ'' galaxies.} Following \cite{Williams2009}, we extracted a sample of galaxies on the basis of the restframe 
color-color diagram defined as
$$(U-V)_{\rm rest}>0.88\times(V-J)_{\rm rest}+0.69\;\;\;[0<z<0.5]$$ 
$$(U-V)_{\rm rest}>0.88\times(V-J)_{\rm rest}+0.59\;\;\;[0.5<z<1],$$
with the additional constraints of $(U-V)_{\rm rest}>1.3$ and $(V-J)_{\rm rest}<1.6$. With these cuts, we selected
4882 galaxies.

\begin{table*}[t!]
\begin{center}
\begin{tabular}{|c|c|c|c|c|c|c|c|c|}
\cline{1-7}
 & & & & & & & & \% w.r.t. \\
 & morphology & red sequence & red UVJ & red SED & quiescent & red \& passive ETGs & & global sample\\
\cline{1-7}
morphology & {\bf 3336} & 77.0\% & 76.9\% & 48.6\% & 56.4\% & 33.0\% & & 19.4\%\\
\cline{1-7}
red sequence & 52.5\% & {\bf 4889} & 83.2\% & 51.4\% & 59.3\% & 31.0\% & & 28.4\%\\
\cline{1-7}
red UVJ & 52.6\% & 83.3\% & {\bf 4882} & 49.9\% & 57.6\% & 30.7\% & & 28.4\%\\
\cline{1-7}
red SED & 62.3\% & 96.6\% & 93.5\% & {\bf 2603} & 84.7\% & 46.3\% & & 15.1\%\\
\cline{1-7}
quiescent & 64.0\% & 98.7\% & 95.8\% & 75.1\% & {\bf 2937} & 47.0\% & & 17.1\%\\
\cline{1-7}
red \& passive ETGs & 72.0\% & 99.2\% & 98.0\% & 78.8\% & 90.2\% & {\bf 1530} & & 8.9\%\\
\cline{1-7}
\end{tabular}
\caption{Passive galaxy samples overlap depending on the different selection criteria adopted and number of galaxies (in boldface). 
Each box gives the percentage of galaxies of a particular sample (specified by the row's name) found in another sample (specified 
by the column's name). The last column also reports the percentage number of each sample with respect to the parent 
zCOSMOS-20k sample of 17211 galaxies.}
\label{tab:tab1}
\end{center}
\end{table*}

\item {\it \bf ``red SED'' galaxies.} \cite{Ilbert2009} employed a set of templates, generated by \cite{Polletta2007} 
with the code GRASIL \citep{Silva1998}, to fit the VIMOS VLT Deep Survey (VVDS) sources \citep{LeFevre2005} 
from the UV-optical to the mid-IR. Therefore, this set of templates provides a better joining of 
UV and mid-IR than those previously proposed by \citep{Ilbert2006}. The nine galaxy templates of \cite{Polletta2007} 
include three SEDs of elliptical galaxies and six templates of spiral galaxies (S0, Sa, Sb, Sc, Sd, Sdm). 
Twelve additional templates obtained from BC03 models with starburst ages ranging from 3 to 0.03 Gyr are also added 
to better represent the data. Finally, the templates were linearly interpolated, so that a total set of 32 templates was obtained. 
These templates were used to estimate photometric types from a best fit to the observed photometry without additional dust extinction. 
With this method, we selected 2603 galaxies best fitted with an early-type template (i.e., SED-type from 1 to 8).
\item {\it \bf ``quiescent'' galaxies.} Following \cite{Ilbert2010} and \cite{Pozzetti2010}, we selected 2937 quiescent galaxies
on the basis of their sSFR, adopting the cut $\rm log(sSFR)<-2\;[Gyr^{-1}]$.
As shown by \cite{Ilbert2010}, this cut corresponds almost directly to a cut ${\rm (NUV-r_{+})_{template}>3.5}$,
with which they defined their sample of quiescent galaxies.
With these selection criteria, we select galaxies that are increasing their mass at a rate less than 1/100th of their
present mass. 
\item {\it \bf ``red \& passive ETGs''.} To obtain a reliable sample with as little bias as possible
due to the presence of star-forming contaminants, we decided to follow the approach used in \cite{Moresco2010}.
From the parent sample, 1530 ETGs were extracted by combining photometric, morphological, 
and optical spectroscopic criteria: galaxies were chosen with a best fit to the SED matching a local E-S0 template \citep[using the 
CWW templates of][]{Ilbert2006}, weak/no emission lines ($\rm{EW_{0}([OII]/H\alpha)}<5$ {\AA}), spheroidal morphology, and an observed $(K-24\mu$m) 
color typical of E/S0 local galaxies (i.e., $(K-24\mu$m)$<-0.5$); for further details about the sample selection, see \cite{Moresco2010}. 
\end{itemize}

Other standard definitions involve, for example, different sets of templates to perform a best-fit match to
the observed SED \citep[CWW photometric types,][]{Ilbert2006} or different cuts in the observed colors,
as for the case of luminous red galaxies \citep[LRG,][]{Eisenstein2001} and Baryon Oscillation Spectroscopic Survey 
\citep[BOSS,][]{Schlegel2009} samples. We checked, however, that the performances of these criteria are much worse
than those of the other samples selected in our analysis in the selection of a pure sample of passive galaxies.
In particular, LRGs and BOSS selections were optimized for the specific properties of those surveys
and cannot straightforwardly be applied to zCOSMOS survey; we refer to \cite{Masters2011} for a discussion
of the contamination by late-type morphological types of BOSS galaxies.

In Fig. \ref{fig:ETGshist} we show the redshift and mass distribution 
of the various samples. Each sample was further divided into six subsamples 
after we considered separately two redshift ranges ($z\leq0.5$ and $z>0.5$), three 
mass ranges (low-mass ${\rm log({\mathcal M}/{\mathcal M}_{\odot})<10.25}$, 
medium-mass ${\rm 10.25\leq log({\mathcal M}/{\mathcal M}_{\odot})<10.75}$, and high-mass 
${\rm log({\mathcal M}/{\mathcal M}_{\odot})\geq10.75}$) in order to be able to discern the dependencies
on mass and redshift. 

In Tab. \ref{tab:tab1} we report the number of galaxies obtained
with the different selections, as well as the percentage overlap
between the various samples. As expected, the overlap between 
samples obtained with different selection criteria is often partial, 
suggesting that they have different properties.
In particular, we notice that there is a small overlap between
all samples and the morphologically selected sample, indicating that
a large fraction of passive galaxies selected on the basis of colors, or more generically on 
the basis of photometric properties, do not have an E/S0 morphology (see Sect. 
\ref{sec:Morpho}). The overlap is also small when considering how many 
morphologically selected ETGs are found in the other samples, indicating 
that a non negligible fraction of ellipticals have blue colors. The overlap is, 
instead, larger when considering the other samples. The red-sequence
and red UVJ samples provide an almost identical number of galaxies
and show a similar reciprocal overlap of $\sim80$\%. However, as we
see in the following section, they have different photometric properties. The fraction of galaxies 
overlapping with the red-sequence sample is above $\sim$80\%, irrespective 
of the selection criterion considered; however this is probably due to the higher 
number of galaxies of that sample. What is more interesting is to consider the 
fraction of galaxies included in the quiescent sample: the overlap with the red SED
and the red \& passive ETGs samples is above $\sim$85\%, which testifies a quiescent star-formation 
activity. However, for the red-sequence and the red UVJ samples, the overlap is $\sim$60\%, indicating 
that almost half of that sample have a higher star-formation activity. The last remark 
is about the red \& passive ETGs criterion, which presents a small overlap with all 
the other samples: this is partly because of the smaller number of galaxies included in this 
sample and partly because this is the only sample also comprising a spectroscopic 
cut (see Sect. \ref{sec:Spectro}).


\section{The analysis}\label{sec:analysis}
As already discussed, the ETGs population is rather homogeneous
in color, star-formation activity, and morphology. We therefore decided to
inspect in detail different properties to study how their global properties 
change for the different samples.

\subsection{Color properties}
\label{sec:col}
The primary observable, apart from the morphology, that we decided to examine, is 
the color. We decided to study both
the standard $(U-B)_{\rm rest}$ color-mass and $(U-V)_{\rm rest}$-$(V-J)_{\rm rest}$
color-color plots, as well as an IRAC color-color diagram as first introduced by \cite{Lacy2004}.

\begin{figure*}[p]
\begin{center}
\includegraphics[angle=-90, width=0.85\textwidth]{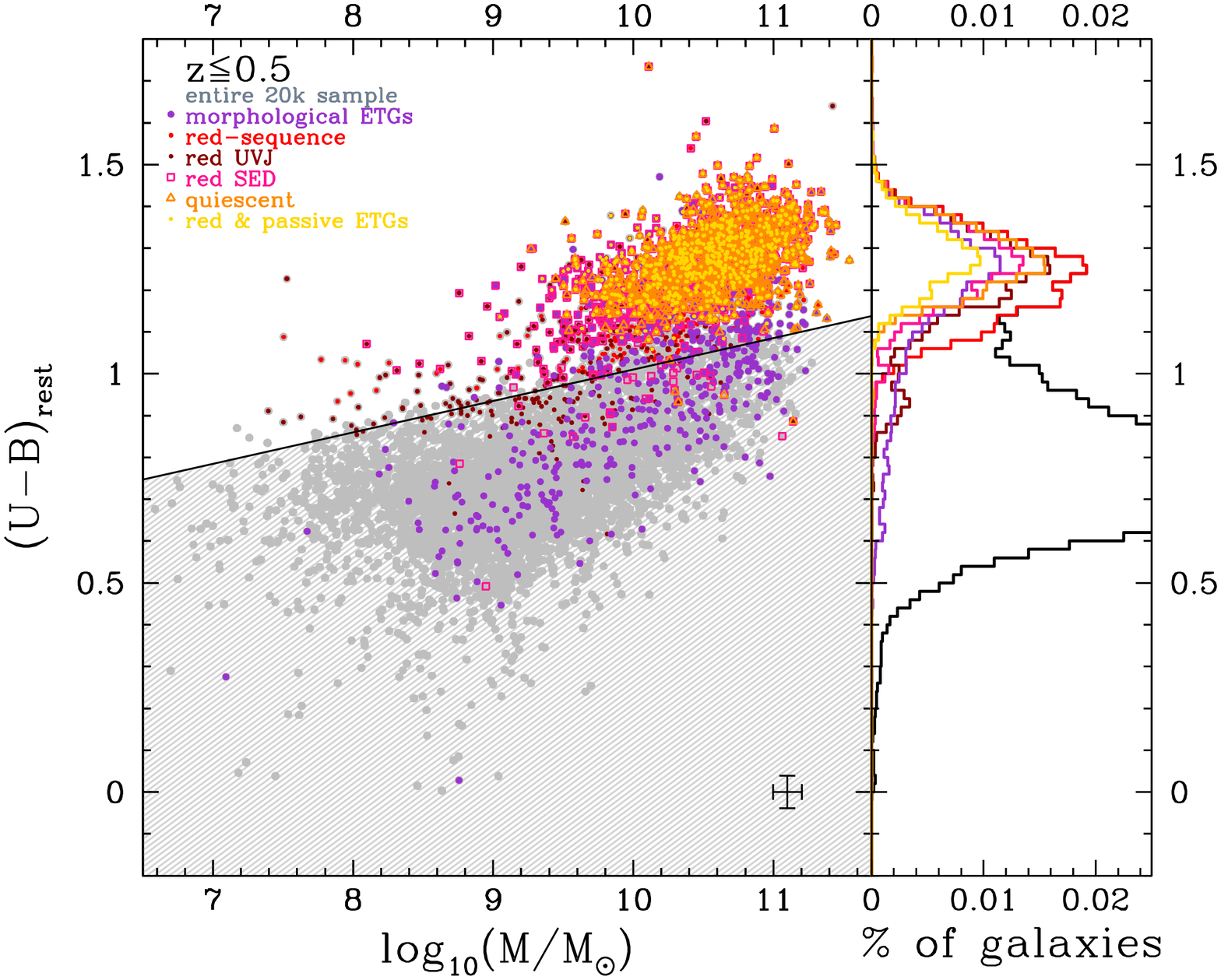}
\includegraphics[angle=-90, width=0.85\textwidth]{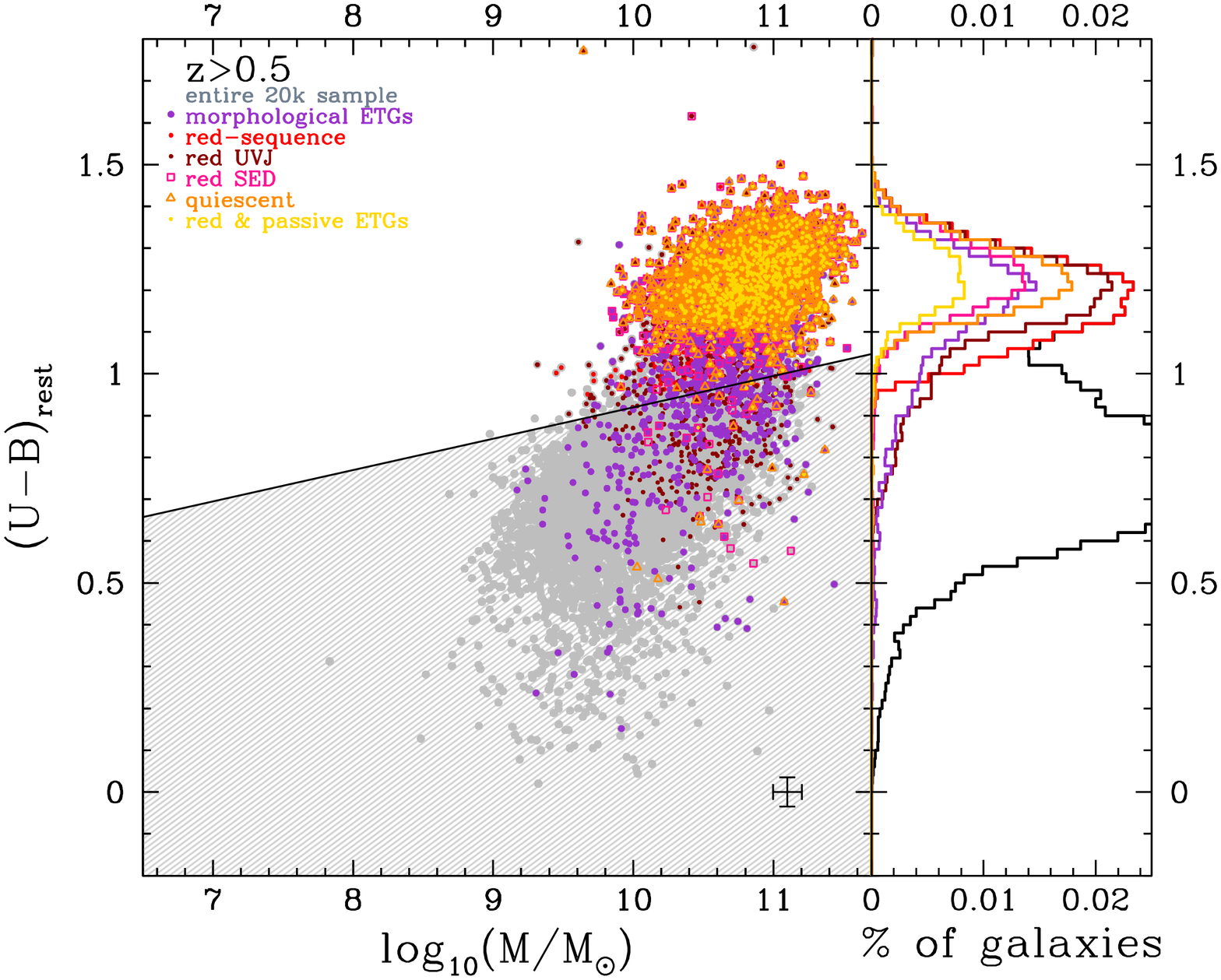}
\end{center}
\caption{$(U-B)_{\rm rest}$-mass diagram. In gray is shown the entire zCOSMOS-20k sample,
in violet the morphological ETGs, in light red the red-sequence galaxies, in dark red the 
red UVJ galaxies, in pink the red SED galaxies, in orange the quiescent galaxies, and in yellow
the red \& passive ETGs. The upper plot shows the $(U-B)_{\rm rest}$-mass 
diagram obtained for $z\leq0.5$, and the lower plot shows the diagram obtained for $z>0.5$.
The gray hatched area represents the region that falls outside the passive region of the 
$(U-B)_{\rm rest}$-mass diagram, as defined in Sect. \ref{sec:criteria}. The representative errorbar
for both quantities is shown in the bottom right corner.
\label{fig:CMD}}
\end{figure*}

The $(U-B)_{\rm rest}$-mass diagram is shown for both redshift ranges ($z\leq0.5$ 
and $z>0.5$) in Fig. \ref{fig:CMD}. The gray hatched area represents the region that
falls outside what is considered the red sequence, as defined by \cite{Peng2010}
and previously reported. The clearest evidence is that most samples lie as expected in the
red-sequence region; this is not surprising, since most selection criteria are based 
(or partially based) on an optical-color selection. For the morphologically selected ETGs, we
found a confirmation of what was deduced from Tab. \ref{tab:tab1}, i.e., that this sample is 
contaminated by the presence of a non-negligible tail of galaxies with blue $(U-B)_{\rm rest}$
colors ($\sim$12-65\%), both at high and low redshifts. Quite surprising is the tail of red UVJ galaxies with blue
color, which even forms a second peak in the $(U-B)_{\rm rest}$ distribution at $z\leq0.5$ (as
shown in Fig. \ref{fig:CMD}). This tail is due to the fact that the cut proposed by 
\cite{Williams2009} does not perfectly fit the observed bimodality in the UVJ diagram of the zCOSMOS-20k sample.

\begin{figure*}[p]
\begin{center}
\includegraphics[angle=-90, width=0.85\textwidth]{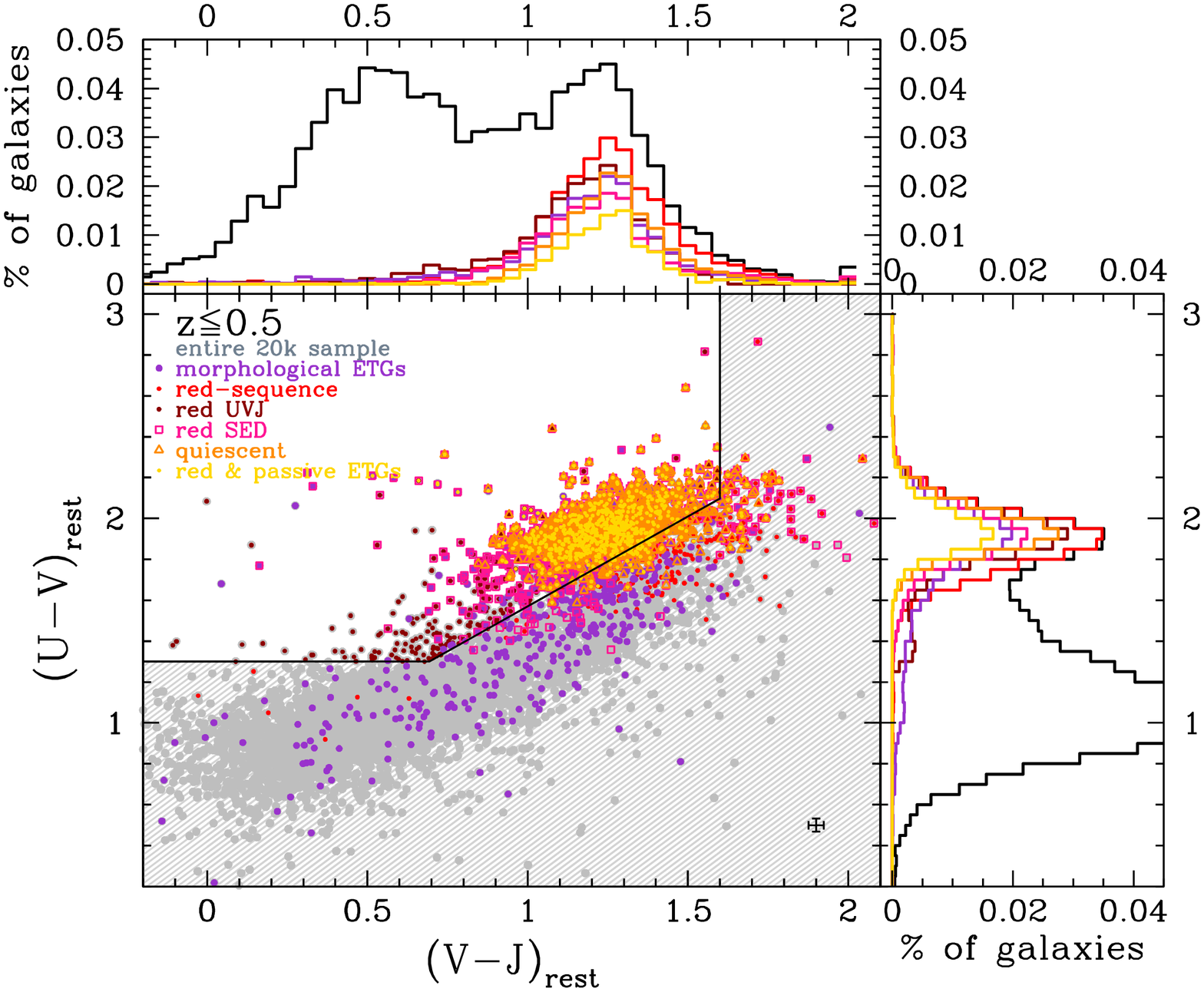}
\includegraphics[angle=-90, width=0.85\textwidth]{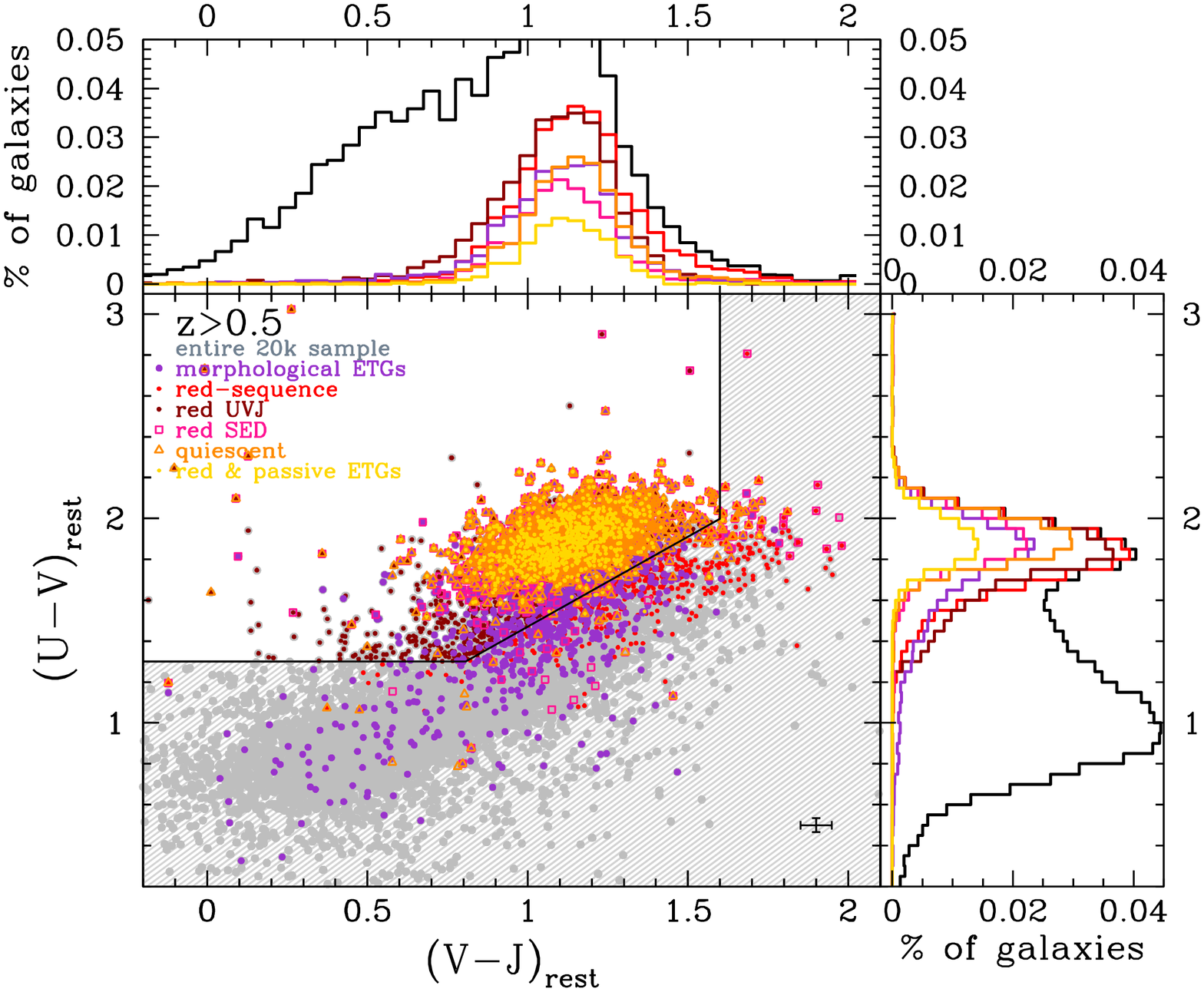}
\end{center}
\caption{$(U-V)_{\rm rest}$-$(V-J)_{\rm rest}$ diagram. In gray is shown the entire zCOSMOS-20k sample,
in violet the morphological ETGs, in light red the red-sequence galaxies, in dark red the 
red UVJ galaxies, in pink the red SED galaxies, in orange the quiescent galaxies, and in yellow
the red \& passive ETGs. The upper plot shows the diagram obtained for $z\leq0.5$, and the 
lower plot for $z>0.5$. The gray hatched area represents the region that falls outside the passive region of the 
UVJ diagram, as defined in Sect. \ref{sec:criteria}. The representative errorbar for both quantities is shown in the bottom right corner.
\label{fig:CCD}}
\end{figure*}

The $(U-V)_{\rm rest}$-$(V-J)_{\rm rest}$ diagram provides complementary information
with respect to the color-mass diagram. As in the previous plot, we notice that the red \& passive ETGs, 
the quiescent galaxies, and the red SED samples well fit the passive region defined by \cite{Williams2009}
and identified by the non-hatched region of Fig. \ref{fig:CCD}; we also observe as before the quite
prominent tail of morphological ETGs with quite blue colors. Less expected is the fact that the
red-sequence galaxies display a non-negligible fraction of galaxies which do not fit perfectly the passive 
UVJ region. We will further discuss the issue of the different behavior of red-sequence and
red UVJ galaxies in Sect. \ref{sec:newcols}.

\begin{figure*}[p]
\begin{center}
\includegraphics[angle=-90, width=0.85\textwidth]{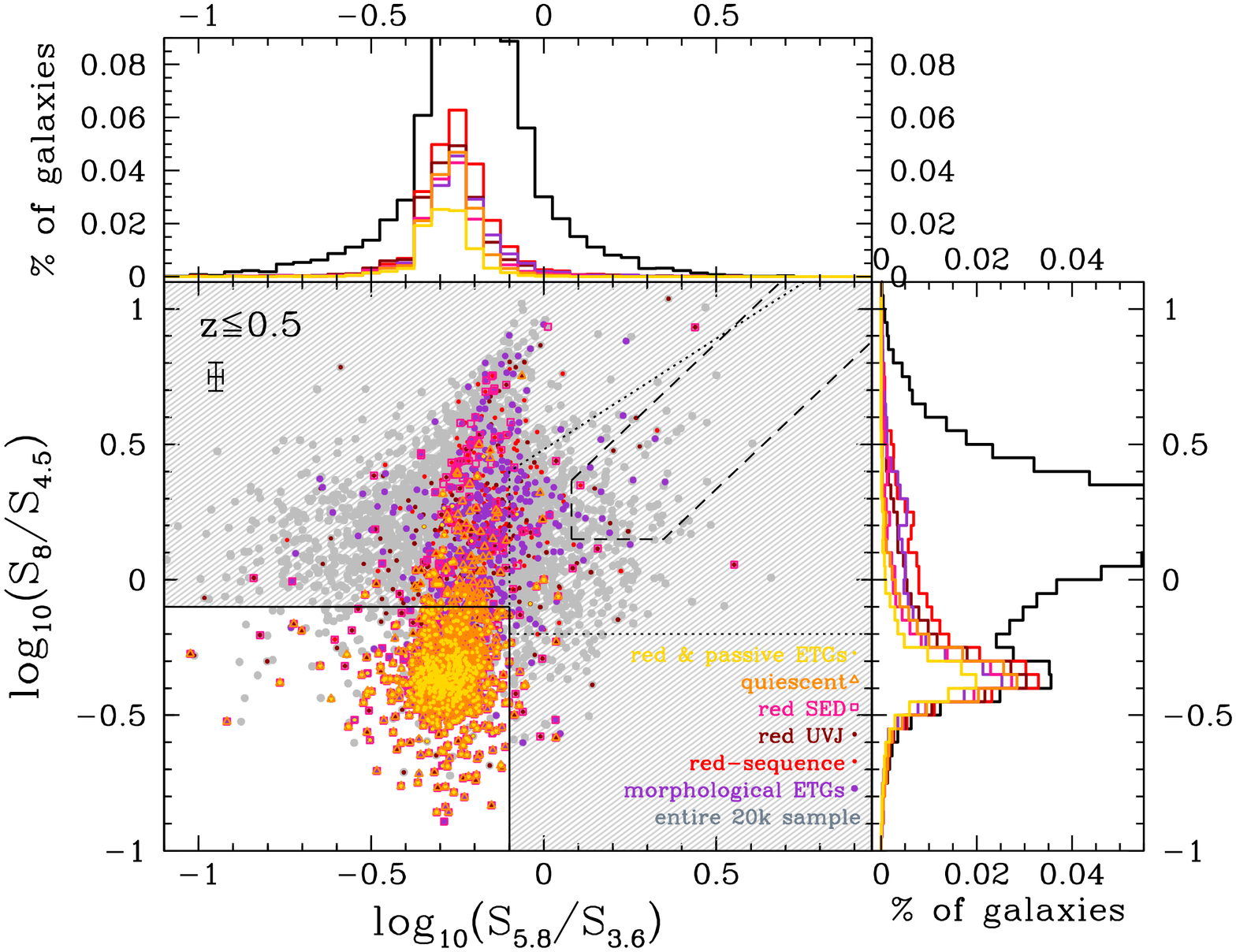}
\includegraphics[angle=-90, width=0.85\textwidth]{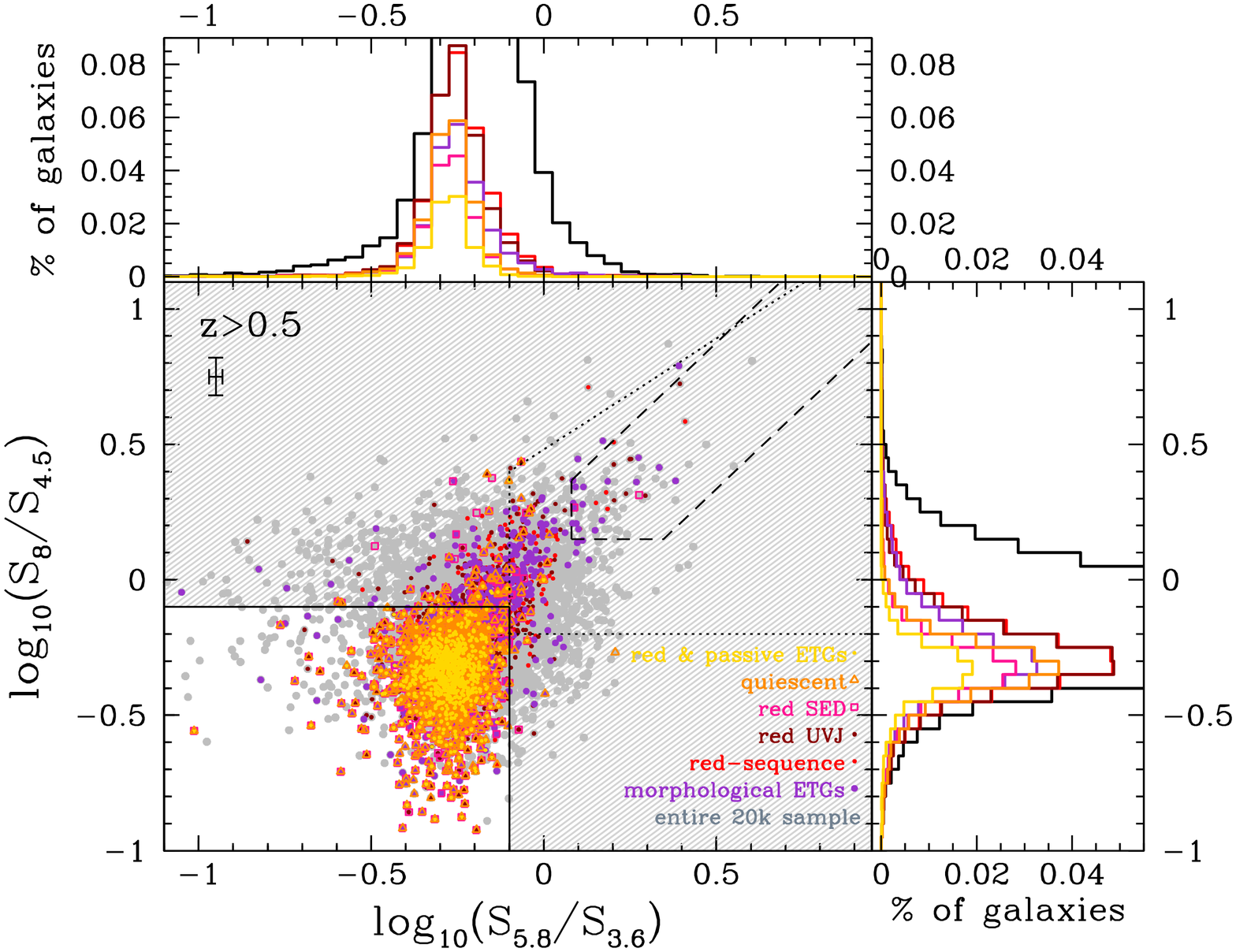}
\end{center}
\caption{$\rm log(S_{8.0}/S_{4.5})-log(S_{5.8}/S_{3.6})$ diagram. In gray is shown the entire zCOSMOS-20k sample,
in violet the morphological ETGs, in light red the red-sequence galaxies, in dark red the 
red UVJ galaxies, in pink the red SED galaxies, in orange the quiescent galaxies, and in yellow
the red \& passive ETGs. The upper plot shows the $\rm log(S_{8.0}/S_{4.5})-log(S_{5.8}/S_{3.6})$ 
diagram obtained for $z\leq0.5$, and the lower plot shows the diagram obtained for $z>0.5$. 
The gray hatched area represents the region that falls outside the passive region of the 
IRAC color-color diagram, as defined in Sect. \ref{sec:col}. The dotted lines represent the region defined
by \cite{Lacy2004} to identify AGN candidates, and the dashed lines the revised definition by \cite{Donley2012}. 
The representative errorbar for both quantities is shown in the bottom right corner.
\label{fig:IRACcol}}
\end{figure*}

The IRAC observed color-color plot is more interesting since these colors less directly (or not at all) influence the
various selection criteria and are therefore a better indication of how much the samples are 
biased. As defined by \cite{Lacy2004}, we plotted the 8.0 $\mu$m/4.5 $\mu$m ratio against the 5.8 
$\mu$m/3.6 $\mu$m ratio in Fig. \ref{fig:IRACcol}. This plot was initially introduced to identify active galactic nuclei (AGN) 
candidates, which should lie in the region identified by the dotted lines. More recently, \cite{Donley2012} 
revised this selection criterion to reduce the contamination by normal star-forming galaxies by narrowing 
the region where the AGN candidates should lie (the area identified by the dashed lines). In their work, they 
also studied the predicted $z=0-3$ IRAC colors of different templates as a function of the AGN fraction, 
considering a star-forming template, a starburst, a normal star-forming spiral galaxy, and an elliptical galaxy. 
By analyzing the tracks of the elliptical galaxies, we found that they occupy a very specific region of the 
IRAC color-color diagram, i.e., $log(S_{8.0}/S_{4.5})<-0.1 \cap log(S_{5.8}/S_{3.6})<-0.1$, at least up to 
$z\sim2$; this region is indicated as the non-hatched region of Fig. \ref{fig:IRACcol}.  From this plot, it is
evident that in both redshift ranges most of the samples are well located in a clump inside the region just 
defined; however, different from the color diagrams, there are more pronounced tails outside this region. 
At $z\leq0.5$ almost all samples show a very significant vertical tail with blue colors in $S_{5.8}/S_{3.6}$ and red colors in 
$S_{8.0}/S_{4.5}$, which is a region in which low-redshift star-forming galaxies lie \citep[for a comparison, 
see Fig. 2 of][]{Donley2012}. In contrast, for $z>0.5$ the tails move toward red colors in both $S_{5.8}/S_{3.6}$ 
and $S_{8.0}/S_{4.5}$. Even if they fall inside the AGN candidate region defined by \cite{Lacy2004}, it is 
very likely that most of these galaxies are higher redshift star-forming galaxies, as can be found following the 
tracks of \cite{Donley2012}. 

\begin{figure*}[p]
\begin{center}
\includegraphics[angle=-90, width=0.85\textwidth]{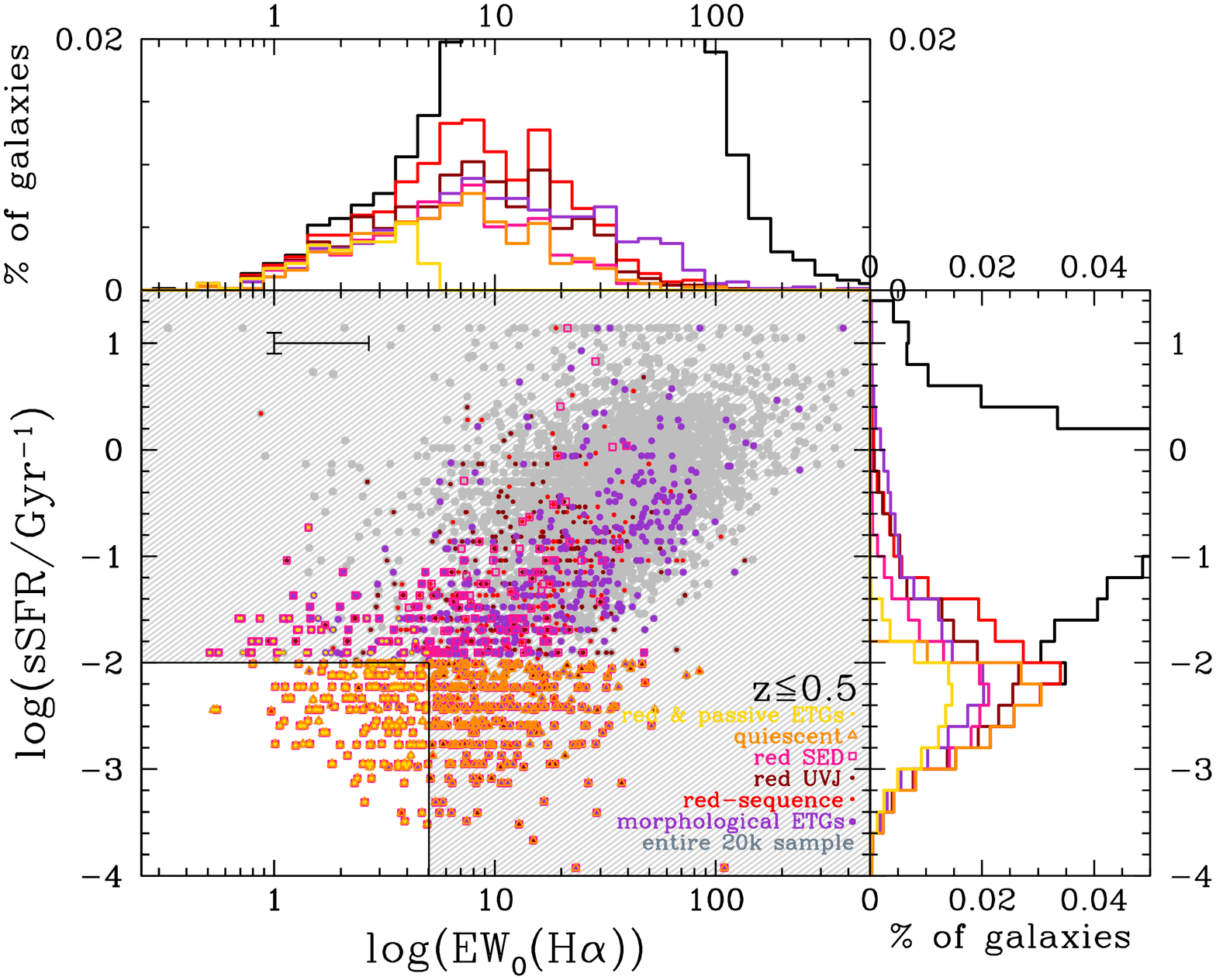}
\includegraphics[angle=-90, width=0.85\textwidth]{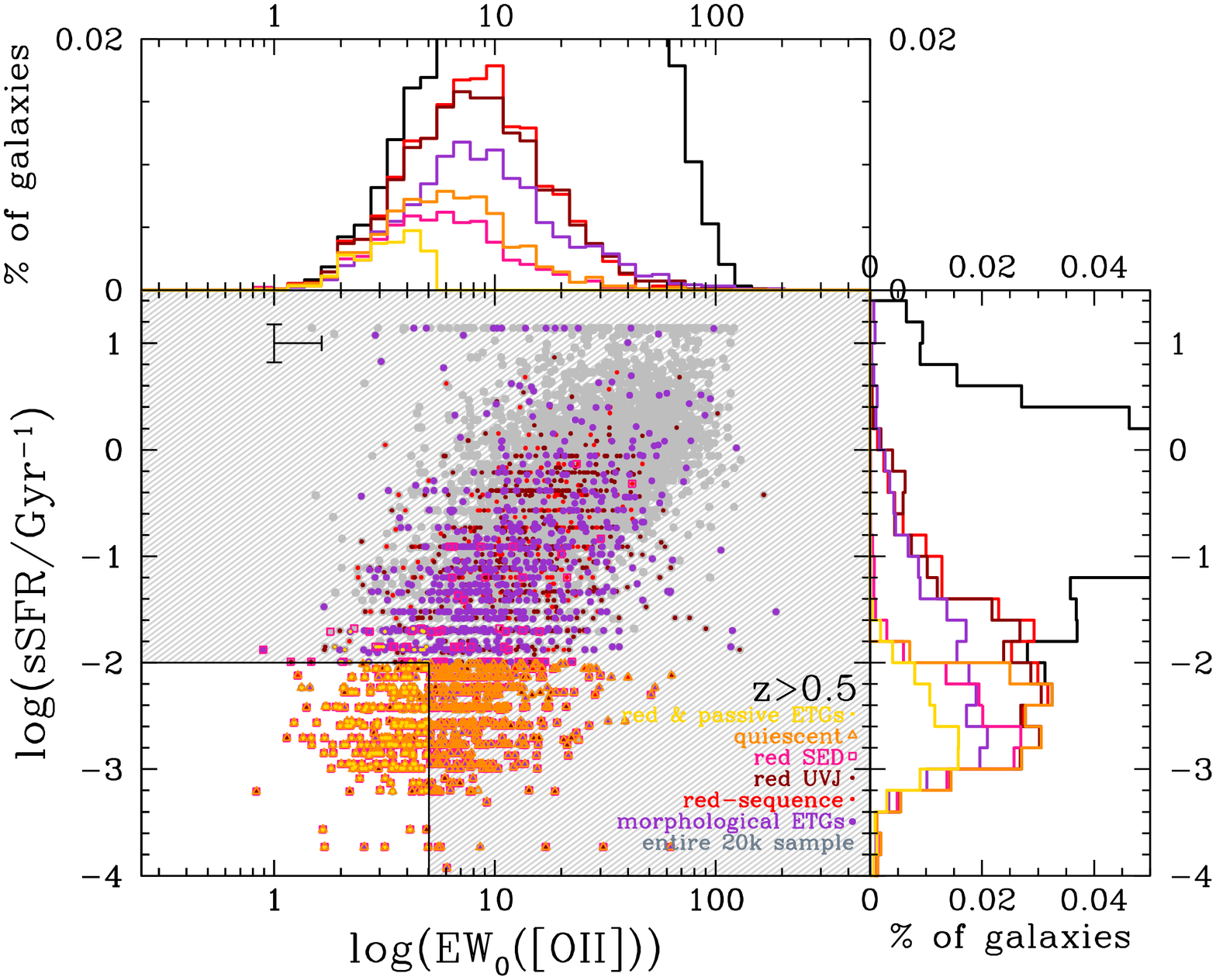}
\end{center}
\caption{sSFR-${\rm EW_{0}([OII]/H\alpha)}$ diagram. In gray is shown the entire zCOSMOS-20k sample,
in violet the morphological ETGs, in light red the red-sequence galaxies, in dark red the 
red UVJ galaxies, in pink the red SED galaxies, in orange the quiescent galaxies, and in yellow
the red \& passive ETGs. The upper plot shows the sSFR-${\rm EW_{0}(H\alpha)}$ 
diagram obtained for $z\leq0.5$, and the lower plot shows the sSFR-${\rm EW_{0}([OII])}$ 
diagram obtained for $z>0.5$. The representative errorbar for both quantities is shown in the upper left corner.
\label{fig:SFRspec}}
\end{figure*}

\begin{figure*}[p]
\begin{center}
\includegraphics[angle=-90, width=0.85\textwidth]{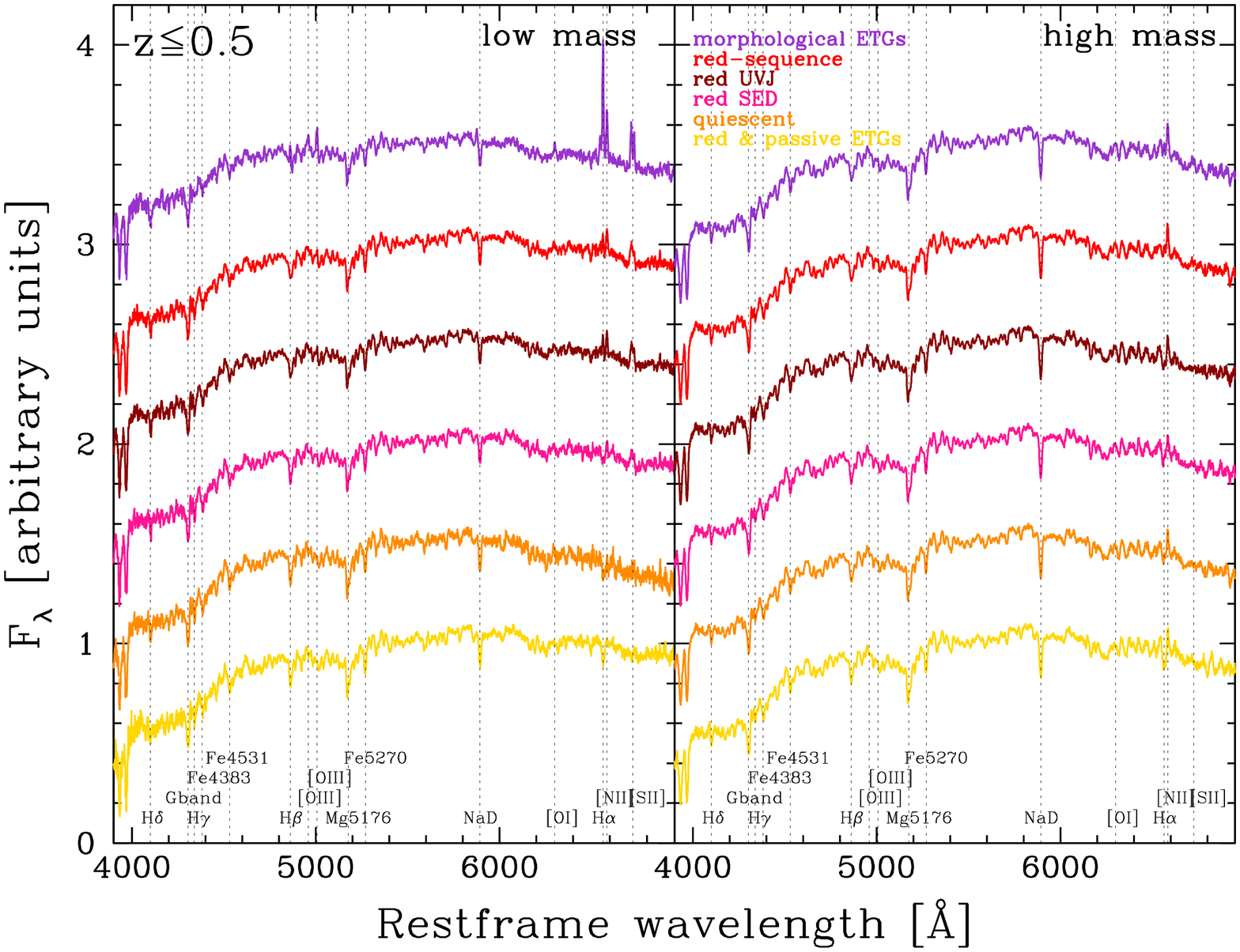}
\includegraphics[angle=-90, width=0.85\textwidth]{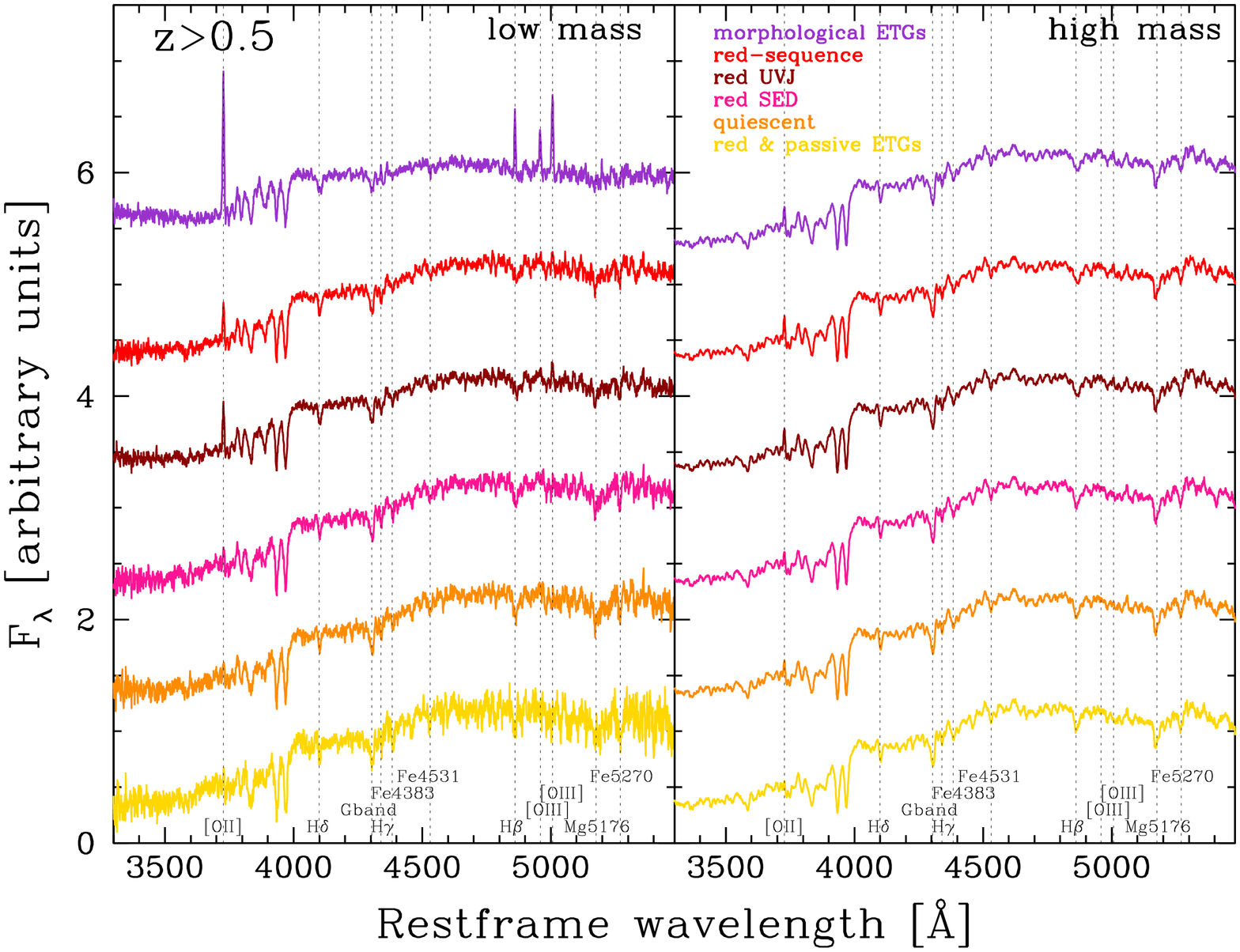}
\end{center}
\caption{Median stacked spectra obtained for $z\leq0.5$ (upper plot) and $z>0.5$ (lower plot); the spectra have
been evaluated in the low-mass bin (${\rm log({\mathcal M}/{\mathcal M}_{\odot})<10.25}$, left panels) and in the high-mass bin 
(${\rm log({\mathcal M}/{\mathcal M}_{\odot})>10.75}$, right panels). In violet are shown the morphological ETGs, in light red the red-sequence galaxies, in dark red the 
red UVJ galaxies, in pink the red SED galaxies, in orange the quiescent galaxies, and in yellow
the red \& passive ETGs.
\label{fig:averagespectra}}
\end{figure*}

\subsection{Spectroscopic properties}
\label{sec:Spectro}
The spectroscopic properties of the galaxies have not been used (except in the red \& passive ETGs criterion)
to select passive galaxies; therefore, they may provide interesting insights concerning the contamination of the various
samples. We decided to look at the restframe equivalent widths ($EW_{0}$) of $\rm [OII]$ and H$\alpha$ lines
since they are well-known indicators of star-formation activity. Given the wavelength coverage of zCOSMOS 
spectra, the two lines are not present in both redshift ranges: in particular H$\alpha$ line is observable at 
$z\leq0.5$, while $\rm [OII]$ is observable at $z>0.5$. These lines will be compared with another indicator of star 
formation, the sSFR, which provides the relative contribution of the SFR by weighting it with 
the total stellar mass of the galaxy.

Figure \ref{fig:SFRspec} shows ${\rm log(sSFR/Gyr^{-1})}$ versus ${\rm EW_{0}(H\alpha)}$ (upper panel; $z>0.5$)
and ${\rm EW_{0}([OII])}$ (lower panel; $z\leq0.5$). As expected, there exists a correlation (even if with a large 
dispersion) between the sSFR and the equivalent widths of [OII] and H$\alpha$ emission lines.
\cite{Ilbert2010}, aiming to study the galaxy stellar mass assembly by morphological and spectral type in the COSMOS
field, identified as quiescent galaxies those with a dereddened color ${\rm(NUV-r^{+})}>3.5$ and consequently having
${\rm log(sSFR/Gyr^{-1})<-2}$. \cite{Mignoli2009}, analyzing the zCOSMOS sample, found that strong and weak 
line emitters can be well divided by an ${\rm EW_{0}([OII])=5}$ {\AA}. Combining this information, we therefore identified 
${\rm log(sSFR/Gyr^{-1})<-2}$ $\cap$ ${\rm EW_{0}(H\alpha\;{\it or }\;[OII])<5}$ {\AA} as the region characteristic of 
quiescent galaxies in this plot (the non-dashed area).

The quiescent galaxies and the red \& passive ETGs samples present, by definition, no tail at high sSFR and EWs, respectively.
From the plot, we find that the color-selected samples (red UVJ and red-sequence) and the morphological
selected sample instead display a marked percentage of galaxies placed at both high sSFR and EWs, both at low and high 
redshifts. While the interpretation of intermediate values of emission lines in terms of star formation
\citep[see for example][]{Yan2006,Yan2012} is not straightforward, especially in the presence of a low level of sSFR, the concomitant higher level of sSFR
clearly indicates a higher level of star formation with respect to the bulk of the population, which lies in the passive region
defined above.

We have also analyzed the stacked median spectra of all samples. In order to disentangle mass and redshift effects,
in Fig. \ref{fig:averagespectra} we plotted only the median spectra for the low- (${\rm log({\mathcal M}/{\mathcal M}_{\odot})<10.25}$) 
and high-mass (${\rm log({\mathcal M}/{\mathcal M}_{\odot})>10.75}$) subsamples, in the low- and high-redshift regimes. 
At both high and low redshift, several emission lines are clearly detectable for most samples. However, one of the most striking result is that
emission lines disappear when passing from the low-mass to the high-mass regime. This represents noticeable evidence that this
population is more quenched than the other. We further inspect this trend in Sect. \ref{sec:resmass}.

The morphological ETGs present the strongest emission lines, especially at low masses. At high redshift, strong emission 
lines in [OII], H$\beta$, and [OIII] are found. At low redshift, the measured ratios between [NII]$\lambda$6583/H$\alpha$ and 
[OIII]$\lambda$5007/H$\beta$ populate the star-forming region in the BPT diagram \citep{Baldwin1981}. This indicates a significant star formation, 
in particular if we consider that we are analyzing median spectra \citep[see also][]{Pozzetti2010}.\\ 
All the other samples show spectra with more typical passive continua, with the presence of faint emission lines in [OII] and 
H$\alpha$ only for the red UVJ and red-sequence galaxy samples.
At high masses, we find no traces of significant emission lines, and all the spectra show features and continua typical
of passively evolving galaxies; this indicates a possible mass dependence that will be further analyzed in Sect. \ref{sec:resmass}. 
In particular, we notice small [OII] emission lines in most spectra at high redshifts, which, however, is compatible with not being 
caused by star-formation activity \citep[see][and the discussion of Sect. \ref{sec:contamination}]{Yan2006,Yan2012}. At low redshift, 
we note that all spectra present a detectable [NII] line (and often also [SII]), but no sign of H$\alpha$, probably due to the fact that the H$\alpha$ emission line is
hidden by the corresponding absorption line.

\subsection{Morphological properties}
\label{sec:Morpho}
\begin{figure}[t!]
\begin{center}
\includegraphics[angle=0, width=0.45\textwidth]{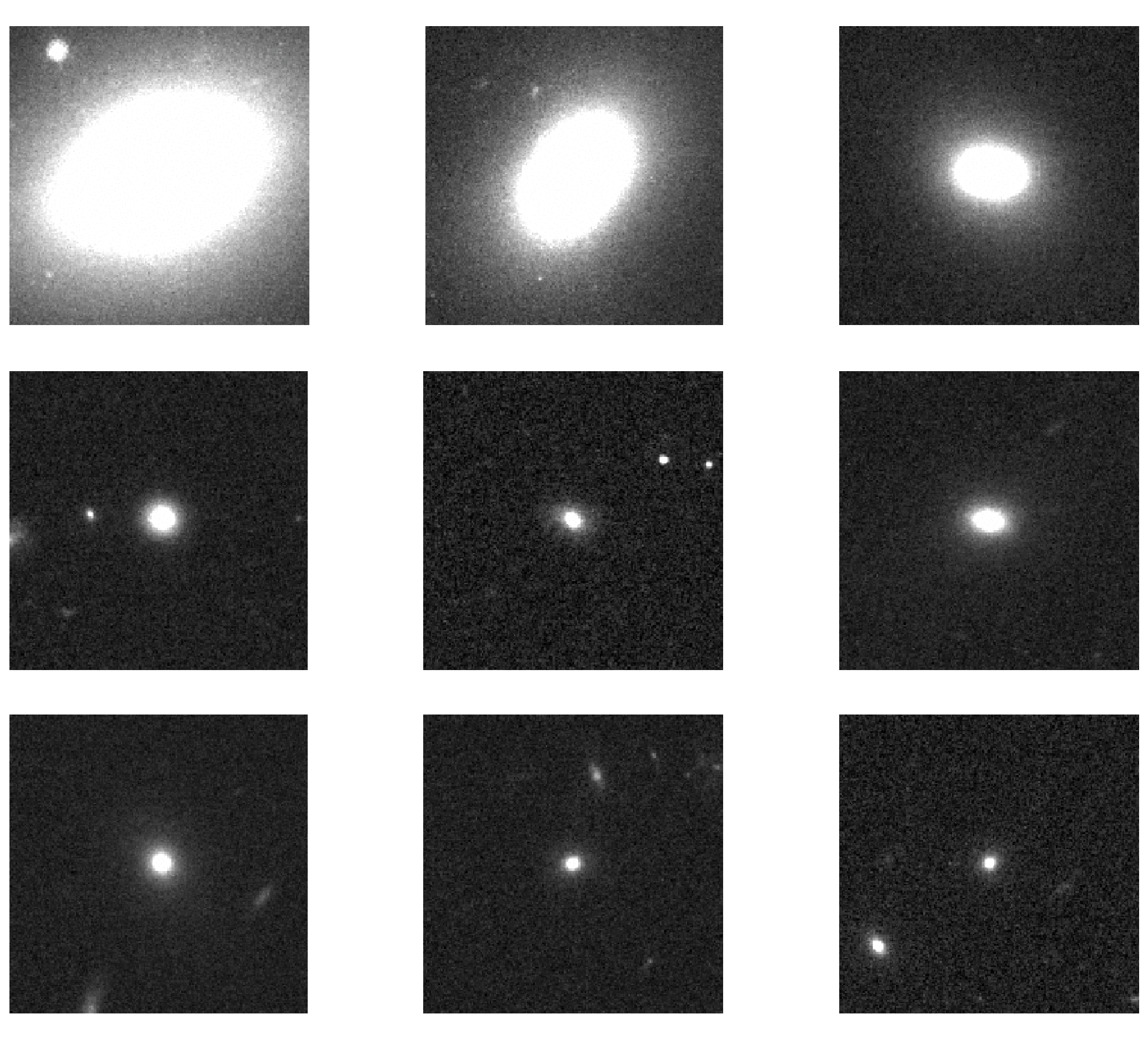}
\end{center}
\caption{Examples of ACS morphology for a subsample of the red \& passive ETGs with elliptical morphologies. The galaxies are shown 
at increasing redshift from upper left to lower right, from $z=0.0791$ to $z=0.9302$.
\label{fig:ETG_morphook}}
\end{figure}

\begin{figure}[t!]
\begin{center}
\includegraphics[angle=0, width=0.45\textwidth]{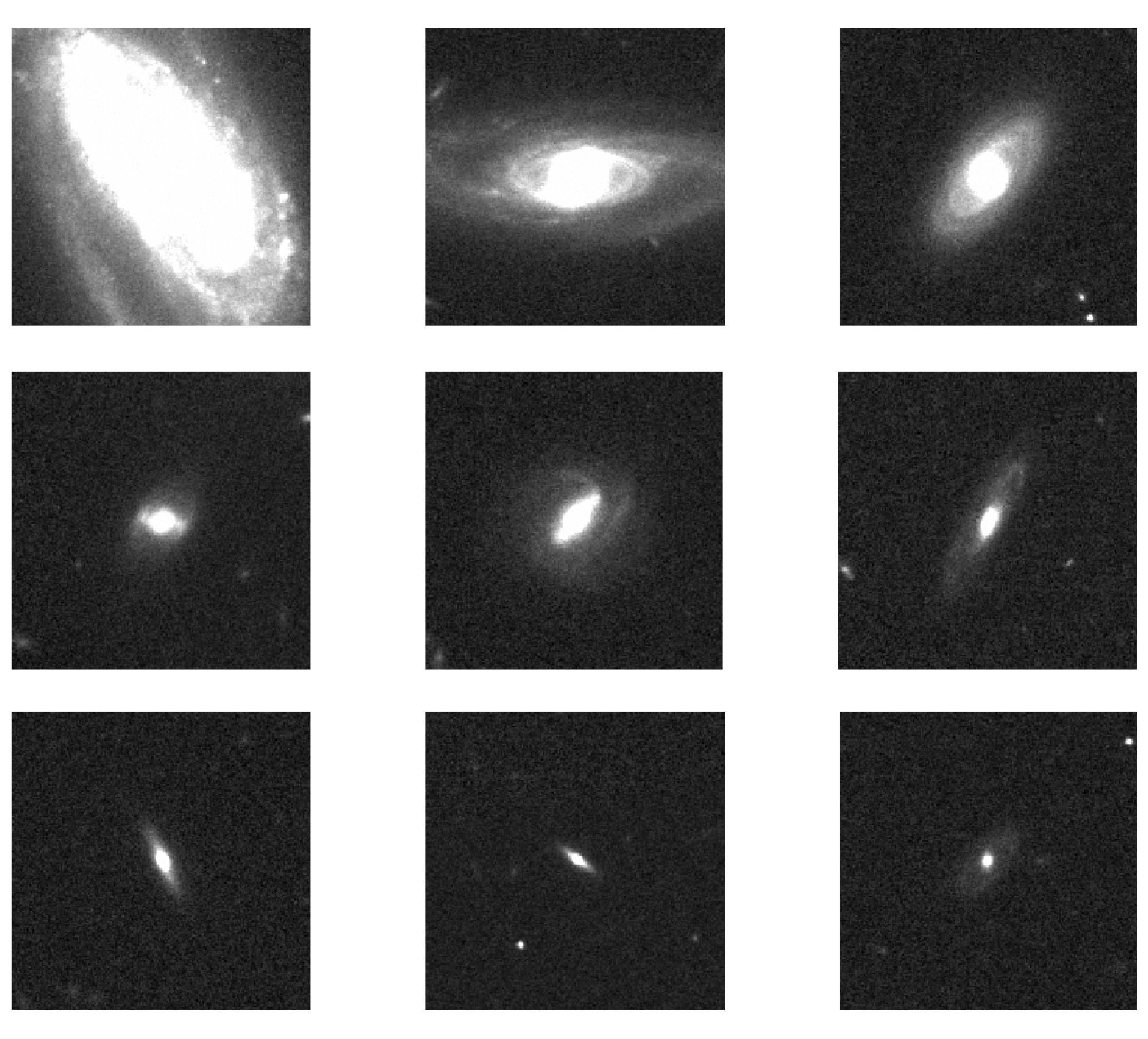}
\end{center}
\caption{Examples of ACS morphology for a subsample of the red \& passive ETGs with very late-type morphologies, found for less than 10\%
of the overall sample. The galaxies are shown at increasing redshift from upper left to lower right, from $z=0.0998$ to $z=0.834$.
\label{fig:ETG_nomorpho}}
\end{figure}

To study in detail the morphological properties, we analyzed the ZEST \citep{Scarlata2007a} types 
of the various samples. While for almost all the samples a large fraction of galaxies have an E/S0 morphology (a clear example
is shown in Fig. \ref{fig:ETG_morphook}), there is a significant fraction, which depends on the adopted selection criteria, of 
galaxies with clearly spiral/irregular morphologies (see Fig. \ref{fig:ETG_nomorpho}). Despite the evident correspondence between 
the bimodality in photometric properties and in morphological types, all the samples are contaminated by late-type morphologies: 
this suggests that the color and the morphological transformation are two distinct processes taking place at different times in the life 
of a galaxies.

This result confirms what was found with independent methods by other analyses \citep[see, for example,][]{Pozzetti2010,Ilbert2010}, i.e., that the morphological
transformation from late-type to early-type galaxies is a process which intervenes on different timescales than
the color transformation. As a result, it is likely for a galaxy, even if it has already stopped its star-formation activity, not to have an
elliptical morphology. In contrast to other properties, it is thus improper to consider those galaxies as star-forming contaminants, 
and the presence of morphologically late-type galaxies will be considered separately.

\section{Results}\label{sec:res}
To quantitatively estimate which is the best (i.e., the least biased as possible) selection criterion,
we estimated the level of contamination by examining different observables, considering colors,
sSFR, spectroscopic features, and visual morphology. Of course, as anticipated, these criteria
should not to be too restrictive because otherwise we trade purity for a poor completeness.

By definition, the less contaminated criterion will be the red \& passive ETGs since it uses all the possible
information available for these galaxies. However, in the context of many galaxy surveys,
it is of utmost interest to also identify a sample that is at the same time the most economic as possible 
(in terms of information used) and as less biased as possible (in terms of star-forming outliers).
To carry on this analysis, it is necessary not only to consider the percentage of contaminants present in each 
sample but also their absolute values, as well as the dependence on the redshift and on the stellar mass of the contamination.

\subsection{Study of the contamination}\label{sec:contamination}
The criteria adopted to define a contaminant are complementary to the ones used to select the various
samples of passive galaxies and are described as:
\begin{itemize}
\item {\it blue $(U-B)_{rest}$ colors}: $$(U-B)_{\rm rest}<1.10+0.075\,log(\frac{\mathcal M}{10^{10}\mathcal M_{\odot}})-0.18\,z$$
\item {\it nonpassive UVJ colors}: $$(U-V)_{\rm rest}<0.88\times(V-J)_{\rm rest}+0.69\;\;\;{\rm[+0.59\;\;for\;\;z>0.5]}$$ 
$$\cup\;\;(U-V)_{\rm rest}<1.3\;\;\cup\;\; (V-J)_{\rm rest}>1.6$$
\item {\it nonpassive IRAC colors}: $$log(S_{8.0}/S_{4.5})>-0.1$$ $$\cup\;\; log(S_{5.8}/S_{3.6})>-0.1$$
\item {\it high sSFR}: $$\rm log(sSFR)>-2\;[Gyr^{-1}]$$
\item {\it presence of emission lines}: 
\begin{center}
$\rm EW_{0}(H\alpha)>5$ {\AA} (if $z\leq0.5$)\\
or $\rm EW_{0}([OII])>5$ {\AA} (if $z>0.5$)
\end{center}
\item {\it nonelliptical morphology}: ZEST type later than E/S0.
\end{itemize}

For each mass and redshift subsample of the various samples defined in Sect. \ref{sec:criteria}, we estimated the
median $(U-B)_{\rm rest}$ color, sSFR, IRAC colors, restframe equivalent width of H$\alpha$ 
([OII] for $z>0.5$), and morphology of the just defined {\it contaminant}, as well as the percentage of 
contamination relative to each subsample. These values are reported in Tab. \ref{tab:tab2}.

As discussed above for the morphology, the spectroscopic properties would also be worth a dedicated discussion,
since it is not straightforward that a weak [OII] or H$\alpha$ emission line is symptomatic of star-formation activity
\citep{Yan2006,Yan2012}. However, for simplicity we address as a contaminant a galaxy having
any of these properties considered non-standard for an early type and examine each one individually.

For each considered property, a trend with stellar mass is evident for all the samples, with a decreasing 
percentage of contaminants with increasing mass. This effect will be analyzed in the following section.
Below we discuss separately the contamination in optical and IRAC colors, in sSFR, in spectroscopy,
and in morphology.

\begin{figure}[t!]
\begin{center}
\includegraphics[angle=0, width=0.47\textwidth]{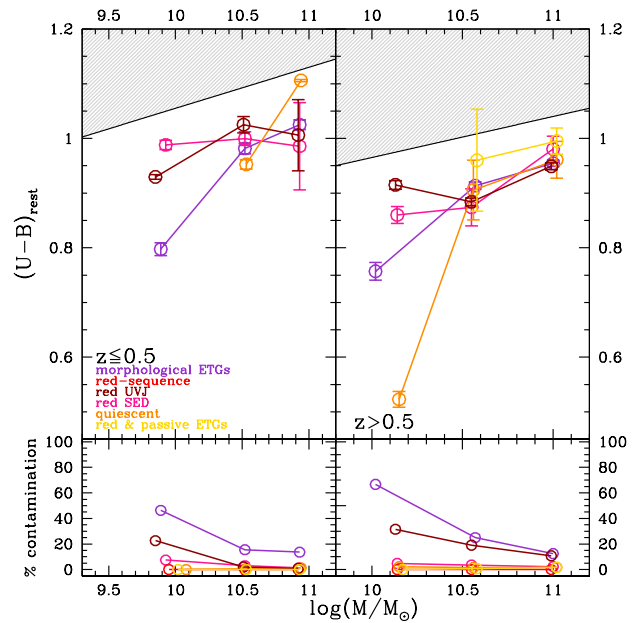}
\end{center}
\caption{Median $(U-B)_{\rm rest}$ color (upper panel) and relative percentage (lower panel) as a function 
of stellar mass of galaxies with blue colors, i.e., $(U-B)_{\rm rest}<1.10+0.075\,log(\frac{\mathcal M}{10^{10}\mathcal M_{\odot}})-0.18\,z$,
for the different selection criteria (in violet the morphological ETGs, in light red the red-sequence galaxies, in dark red the 
red UVJ galaxies, in pink the red SED galaxies, in orange the quiescent galaxies, and in yellow
the red \& passive ETGs). The errorbars represent the error on the median.
\label{fig:contamination1}}
\end{figure}

\paragraph{\bf Blue $(U-B)_{\rm rest}$ colors.} In Sect. \ref{sec:col} we have shown the $(U-B)_{\rm rest}$-mass diagram for all the samples,
proving that, given the adopted selection criteria, almost all the selected galaxies lie in the red sequence, with minor tails extending 
in the blue cloud. As a consequence, the number of outliers with clearly blue colors are a minor percentage, typically $\lesssim$5\%
for all criteria (except morphological and red UVJ galaxies) at both low and high redshift. Moreover, as can be inferred
from Tab. \ref{tab:tab2}, this small fraction of blue outliers has relatively red colors, very close to the considered separation
between red sequence and blue cloud. In contrast, morphological ETGs display a substantial percentage of galaxies,
$\sim12-65\%$ depending on the redshift range and on the mass regime, with a median $(U-B)_{\rm rest}$ color bluer
with respect to the contaminants present in the other samples. This fact testifies that even if a correlation exists between
photometric properties and morphology of passive galaxies, it is not a one-to-one correlation, and a considerable number
of blue ellipticals coexist with their more standard red counterparts \citep[see also][]{Tasca2009}. Also the red UVJ sample
presents a significant contamination, $\sim$30-10\% depending on the redshift range and on the mass regime. This is probably
due to the fact that the color cuts defined by \cite{Williams2009} do not seem to reproduce well the observed bimodality
in the UVJ diagram (as it can be seen by Fig. \ref{fig:CCD}), with tails evident in the blue part of the $(U-V)_{\rm rest}$
histogram. However, the median color of the contaminant is, as in most of the previous cases, very close to the adopted
definition of red sequence. All these data are reported in Tab. \ref{tab:tab2} and shown in Fig. \ref{fig:contamination1}.

\paragraph{\bf Nonpassive UVJ colors.} For the UVJ color-color diagram, we considered as a contaminant a galaxy with $(U-V)_{\rm rest}$
and $(V-J)_{\rm rest}$ colors outside the region defined by \cite{Williams2009}. Therefore we just estimated the percentage of these galaxies 
for each selection criterion; we refer to Sect. \ref{sec:col} for the discussion about where those contaminants lie in the UVJ color-color diagram.

This analysis confirms the results just discussed for the color-mass diagram: the quiescent galaxies, red SED galaxies, and red \& passive ETGs fit almost perfectly to the passive region
in the UVJ diagram, with a percentage of contaminants always below 20\% at low redshift and below 5\% at high redshift.
The percentage of contaminants is slightly higher for the red-sequence sample, around 10-30\%, and this fact demonstrates that
the cut defined by \cite{Peng2010} in the $(U-B)_{\rm rest}$ is also not optimal in dividing passive and star-forming galaxies.
As in the previous analysis, the morphological ETGs are the sample with the highest contamination in UVJ colors, around
12-65\% depending on the redshift and mass ranges.

\begin{figure}[t!]
\begin{center}
\includegraphics[angle=0, width=0.47\textwidth]{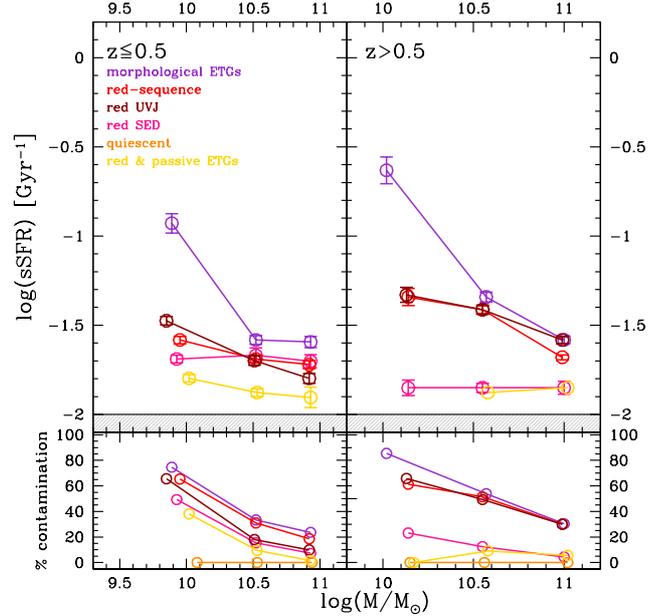}
\end{center}
\caption{Median $\rm log(sSFR/Gyr^{-1})$ (upper panel) and relative percentage (lower panel) as a function 
of stellar mass of galaxies with high sSFR, i.e., $\rm log(sSFR/Gyr^{-1})>-2$, for the different selection criteria 
(in violet the morphological ETGs, in light red the red-sequence galaxies, in dark red the 
red UVJ galaxies, in pink the red SED galaxies, in orange the quiescent galaxies, and in yellow
the red \& passive ETGs). The errorbars represent the error on the median.
\label{fig:contamination2}}
\end{figure}

\paragraph{\bf Nonpassive IRAC colors.} The study of the IRAC colors allows a clearer estimate of the presence of star-forming 
contaminants since these bands were not directly used to define ETG samples (they were used in the red SED galaxies, 
but together with all the other photometric bands for the best fit to the SEDs).
We considered as a contaminant a galaxy falling outside the passive IRAC selection defined in Sect. \ref{sec:col}.
Therefore, since it is possible for such galaxy to have a redder $S_{5.8}/S_{3.6}$ color or a redder $S_{8.0}/S_{4.5}$ color, or both,
as for the UVJ case, we just estimated the percentage of these galaxies for each selection criterion and refer to Sect. \ref{sec:col} for the discussion about
where those contaminants lie in the IRAC color-color diagram.

The morphological ETGs are the sample with the higher percentage of contamination, $\sim$20-75\% depending on the mass 
and redshift range. Similar to the case of the sSFR, the red UVJ and the red-sequence galaxies are more 
biased, with $\sim$25-55\% (depending on the mass range) of the sample not classified as passive for the IRAC color-color 
criterion at $z\leq0.5$ and $\sim$10-50\% at $z>0.5$. Apart from the red \& passive ETGs sample, with a contamination of $\sim5-25$\% at low redshift
and of $\sim 2-40$\% at high redshift (from the lowest to the highest masses), the purest samples are the quiescent and the red SED
galaxies ($\sim10-45$\% at low redshift and $\sim3-45$\% at high redshift).

\begin{figure}[t!]
\begin{center}
\includegraphics[angle=0, width=0.47\textwidth]{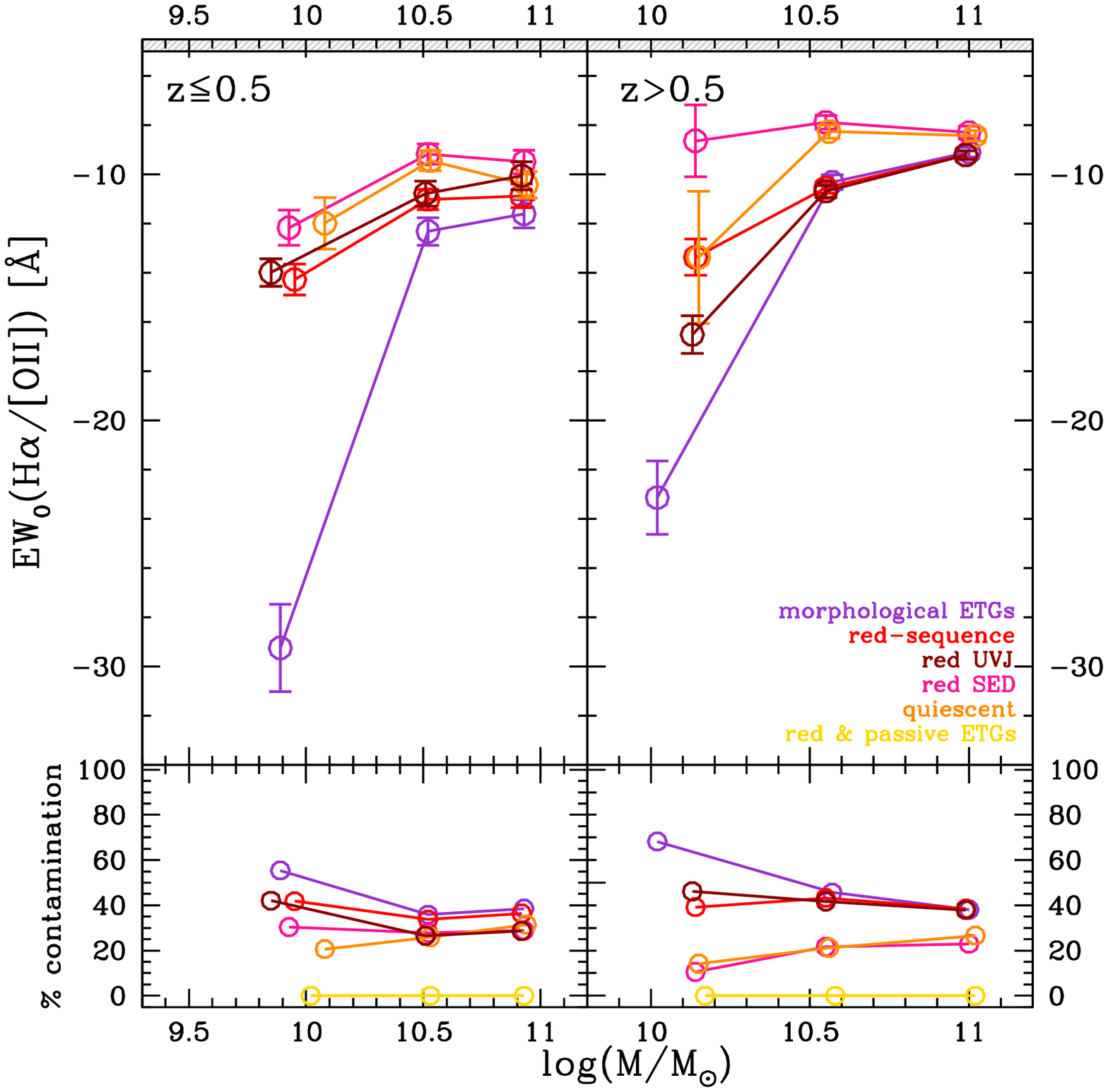}
\end{center}
\caption{Median $\rm EW_{0}(H\alpha)$ (upper panel; $\rm EW_{0}([OII])$ if $z>0.5$) and relative percentage 
(lower panel) as a function of stellar mass of galaxies with significant emission lines, i.e., $\rm EW_{0}(H\alpha)>5$ 
{\AA} (if $z\leq0.5$) and $\rm EW_{0}([OII])>5$ {\AA} (if $z>0.5$), for the different selection criteria (in violet the 
morphological ETGs, in light red the red-sequence galaxies, in dark red the red UVJ galaxies, in pink the 
red SED galaxies, in orange the quiescent galaxies, and in yellow the red \& passive ETGs).
The errorbars represent the error on the median. By convention, the emission lines are quoted with negative values.
\label{fig:contamination3}}
\end{figure}

\paragraph{\bf High sSFR.} Considering the sSFR, the morphological ETGs are the sample with
the most extended tails ($\sim$30-90\%), with median values up to ${\rm log(sSFR/Gyr^{-1})\sim-0.6}$, corresponding to a star-formation activity
$\sim$25 times higher with respect to the adopted definition of quiescent (i.e., ${\rm log(sSFR/Gyr^{-1})<-2}$).
Both red UVJ and red-sequence samples represent an intermediate case, with a median ${\rm log(sSFR/Gyr^{-1})\sim-1.5/-1.7}$
for the contaminants ($\sim$3 times more star-formation activity than the assumed quiescent limit), and a percentage of contamination
$\sim$20-65\%, depending on the mass range.
The less biased samples (the quiescent sample is by definition not biased with respect to this parameter) are the red SED 
and the red \& passive ETGs ones, having median values of sSFR much closer to the quiescent limit and a percentage of contamination
in most cases $\lesssim$15\%.
All these data are reported in Tab. \ref{tab:tab2} and shown in Fig. \ref{fig:contamination2}.

\begin{table*}[t!]
\begin{center}
\begin{tabular}{|c|cc|cc|cc|cc|cc|cc|}
\hline
 & \multicolumn{6}{c|} {$z\leq0.5$} & \multicolumn{6}{c|} {$z>0.5$}\\
 & \multicolumn{2}{c}{low-mass} & \multicolumn{2}{c}{med-mass} & \multicolumn{2}{c|}{high-mass} & \multicolumn{2}{c}{low-mass} & \multicolumn{2}{c}{med-mass} & \multicolumn{2}{c|}{high-mass}\\
 \cline{2-13}
 & {\tiny median} & {\tiny \%} & {\tiny median} & {\tiny \%} & {\tiny median} & {\tiny \%} & {\tiny median} & {\tiny \%} & {\tiny median} & {\tiny \%} & {\tiny median} & {\tiny \%} \\
\hline
{\bf blue $(U-B)_{\rm rest}$} & & & & & & & & & & & & \\
{\bf colors} & & & & & & & & & & & & \\
\cline{1-1}
morphology & 0.80 & 46.3\% & 0.98 & 15.5\% & 1.03 & 13.6\% & 0.76 & 66.7\% & 0.91 & 25\% & 0.95 & 12.4\% \\
red-sequence & -- & -- & -- & -- & -- & -- & -- & -- & -- & -- & -- & -- \\
red UVJ & 0.93 & 22.6\% & 1.03 & 2\% & 1.01 & 0.9\% & 0.91 & 31.5\% & 0.88 & 19.2\% & 0.95 & 10.8\% \\
red SED & 0.99 & 7.5\% & 1.00 & 3\% & 0.98 & 1.3\% & 0.86 & 4.8\% & 0.87 & 3.6\% & 0.98 & 2.3\% \\
passive & -- & 0.0\% & 0.95 & 0.8\% & 1.11 & 1.3\% & 0.52 & 2.2\% & 0.91 & 1.6\% & 0.96 & 1.7\% \\
red \& passive ETGs & -- & 0.0\% & -- & 0.0\% & -- & 0.0\% & -- & 0.0\% & 0.96 & 0.7\% & 1 & 1.8\% \\
\hline
{\bf nonpassive} & & & & & & & & & & & & \\
{\bf UVJ colors} & & & & & & & & & & & & \\
\cline{1-1}
morphology & \multicolumn{2}{c|}{46.6\%} & \multicolumn{2}{c|}{30.4\%} & \multicolumn{2}{c|}{32.7\%} & \multicolumn{2}{c|}{63\%} & \multicolumn{2}{c|}{25.7\%} & \multicolumn{2}{c|}{12.1\%} \\
red-sequence & \multicolumn{2}{c|}{23.1\%} & \multicolumn{2}{c|}{28.2\%} & \multicolumn{2}{c|}{29.5\%} & \multicolumn{2}{c|}{23.3\%} & \multicolumn{2}{c|}{24.3\%} & \multicolumn{2}{c|}{12.7\%} \\
red UVJ & \multicolumn{2}{c|}{--} & \multicolumn{2}{c|}{--} & \multicolumn{2}{c|}{--} & \multicolumn{2}{c|}{--} & \multicolumn{2}{c|}{--} & \multicolumn{2}{c|}{--} \\
red SED & \multicolumn{2}{c|}{8.9\%} & \multicolumn{2}{c|}{16.5\%} & \multicolumn{2}{c|}{18.2\%} & \multicolumn{2}{c|}{5.8\%} & \multicolumn{2}{c|}{5.6\%} & \multicolumn{2}{c|}{4.6\%} \\
quiescent & \multicolumn{2}{c|}{1.4\%} & \multicolumn{2}{c|}{13.6\%} & \multicolumn{2}{c|}{22.2\%} & \multicolumn{2}{c|}{3.3\%} & \multicolumn{2}{c|}{3.7\%} & \multicolumn{2}{c|}{3.6\%} \\
red \& passive ETGs & \multicolumn{2}{c|}{3.9\%} & \multicolumn{2}{c|}{13.2\%} & \multicolumn{2}{c|}{10.8\%} & \multicolumn{2}{c|}{0.0\%} & \multicolumn{2}{c|}{1.1\%} & \multicolumn{2}{c|}{1.5\%} \\
\hline
{\bf nonpassive} & & & & & & & & & & & & \\
{\bf IRAC colors} & & & & & & & & & & & & \\
\cline{1-1}
morphology & \multicolumn{2}{c|}{60.1\%} & \multicolumn{2}{c|}{26.6\%} & \multicolumn{2}{c|}{26.0\%} & \multicolumn{2}{c|}{60.4\%} & \multicolumn{2}{c|}{23.3\%} & \multicolumn{2}{c|}{7.6\%} \\
red-sequence & \multicolumn{2}{c|}{55.3\%} & \multicolumn{2}{c|}{26.1\%} & \multicolumn{2}{c|}{25.1\%} & \multicolumn{2}{c|}{51.3\%} & \multicolumn{2}{c|}{21.6\%} & \multicolumn{2}{c|}{8.4\%} \\
red UVJ & \multicolumn{2}{c|}{55.2\%} & \multicolumn{2}{c|}{13.6\%} & \multicolumn{2}{c|}{9.6\%} & \multicolumn{2}{c|}{49.6\%} & \multicolumn{2}{c|}{18.5\%} & \multicolumn{2}{c|}{7.2\%} \\
red SED & \multicolumn{2}{c|}{44.7\%} & \multicolumn{2}{c|}{14.4\%} & \multicolumn{2}{c|}{11\%} & \multicolumn{2}{c|}{41.4\%} & \multicolumn{2}{c|}{10.9\%} & \multicolumn{2}{c|}{2.6\%} \\
quiescent & \multicolumn{2}{c|}{29.6\%} & \multicolumn{2}{c|}{8.6\%} & \multicolumn{2}{c|}{14.4\%} & \multicolumn{2}{c|}{43.5\%} & \multicolumn{2}{c|}{9.4\%} & \multicolumn{2}{c|}{2.5\%} \\
red \& passive ETGs & \multicolumn{2}{c|}{25.2\%} & \multicolumn{2}{c|}{7\%} & \multicolumn{2}{c|}{5.2\%} & \multicolumn{2}{c|}{38.9\%} & \multicolumn{2}{c|}{3.2\%} & \multicolumn{2}{c|}{1.6\%} \\
\hline
{\bf high sSFR} & & & & & & & & & & & & \\
\cline{1-1}
morphology & -0.93 & 74.6\% & -1.58 & 33.3\% & -1.59 & 23.7\% & -0.63 & 85.4\% & -1.34 & 53.8\% & -1.58 & 30.3\% \\
red-sequence & -1.58 & 65.4\% & -1.69 & 31.1\% & -1.72 & 18.7\% & -1.34 & 61.2\% & -1.41 & 51.4\% & -1.68 & 30.3\% \\
red UVJ & -1.48 & 65.6\% & -1.70 & 18\% & -1.80 & 9.9\% & -1.3 & 65.8\% & -1.41 & 49.3\% & -1.58 & 30.1\% \\
red SED & -1.69 & 49.3\% & -1.67 & 15.4\% & -1.70 & 7.2\% & -1.85 & 12.4\% & -1.85 & 13.2\% & -1.85 & 4.4\% \\
quiescent & -- & -- & -- & -- & -- & -- & -- & -- & -- & -- & -- & -- \\
red \& passive ETGs & -1.80 & 38.1\% & -1.88 & 9.5\% & -1.91 & 1.6\% & -- & 0.0\% & -1.88 & 9.1\% & -1.85 & 5.6\% \\
\hline
{\bf presence of} & & & & & & & & & & & & \\
{\bf emission lines} & & & & & & & & & & & & \\
\cline{1-1}
morphology & -29.2 & 55.4\% & -12.3 & 35.9\% & -11.6 & 38.6\% & -23.1 & 68.2\% & -10.3 & 45.6\% & -9.1 & 38\% \\
red-sequence & -14.3 & 41.8\% & -11 & 33.8\% & -10.9 & 36.3\% & -13.4 & 39.2\% & -10.5 & 43.3\% & -9.2 & 38.5\% \\
red UVJ & -14.0 & 42.1\% & -10.8 & 26.6\% & -10.1 & 28.7\% & -16.5 & 46.2\% & -10.7 & 41.7\% & -9.2 & 37.9\% \\
red SED & -12.2 & 30.4\% & -9.2 & 27.8\% & -9.5 & 28.9\% & -8.6 & 10.75\% & -7.9 & 21.6\% & -8.3 & 23\% \\
quiescent & -12.0 & 20.7\% & -9.4 & 25.9\% & -10.4 & 31.4\% & -13.4 & 14.1\% & -8.3 & 21.2\% & -8.4 & 26.6\% \\
red \& passive ETGs & -- & -- & -- & -- & -- & -- & -- & -- & -- & -- & -- & -- \\
\hline
{\bf nonelliptical} & & & & & & & & & & & & \\
{\bf morphology} & & & & & & & & & & & & \\
\cline{1-1}
morphology & -- & -- & -- & -- & -- & -- & -- & -- & -- & -- & -- & -- \\
red-sequence & 2.1 & 60.8\% & 2.1 & 39\% & 2.1 & 22.8\% & 2.1 & 67.2\% & 2.1 & 55.4\% & 2.1 & 31.5\% \\
red UVJ & 2.1 & 60.6\% & 2.1 & 31\% & 2.1 & 15.2\% & 2.1 & 67.3\% & 2.1 & 52.7\% & 2.1 & 29.9\% \\
red SED & 2.1 & 51.9\% & 2.1 & 30\% & 2.1 & 18.9\% & 2.1 & 59.6\% & 2.1 & 44.8\% & 2.1 & 22.9\% \\
quiescent & 2.1 & 50.2\% & 2.1 & 31.4\% & 2.1 & 18.6\% & 2.1 & 63\% & 2.1 & 45.7\% & 2.1 & 23.7\% \\
red \& passive ETGs & 2.1 & 34.8\% & 2.1 & 25.7\% & 2.1 & 11.3\% & 2.1 & 77.8\% & 2.1 & 40.7\% & 2.1 & 15.1\% \\
\hline
\end{tabular}
\caption{Contamination of the different samples of passive galaxies as a function of stellar mass and redshift. The table reports the median values 
of colors, sSFR, equivalent widths and morphology for the galaxies having blue colors, high specific star formation rates, significant 
emission lines and late-type morphologies as defined in Sect. \ref{sec:contamination}, and the percentage relative to each subsample.
The low-mass bin refers to ${\rm log({\mathcal M}/{\mathcal M}_{\odot})<10.25}$, the med-mass bin to 
${\rm 10.25<log({\mathcal M}/{\mathcal M}_{\odot})<10.75}$, and the high-mass bin to ${\rm log({\mathcal M}/{\mathcal M}_{\odot})>10.75}$.
As defined in the text, the emission lines reported are the median restframe equivalent widths (in units of [\AA]) of H$\alpha$ for $z\leq0.5$
and of [OII] for $z>0.5$; by convention, the emission lines are quoted with negative values.}
\label{tab:tab2}
\end{center}
\end{table*}

\paragraph{\bf Presence of emission lines.} 
We estimated the median equivalent width of H$\alpha$ at $z\leq0.5$, and [OII] at $z>0.5$ for
the galaxies with significant emission lines (i.e., $>$5 {\AA}) in each sample; the results are
reported in Tab \ref{tab:tab2} and shown in Fig. \ref{fig:contamination3}. The sample with the
strongest emission lines is the morphological ETGs, with median equivalent widths up to 
$\sim$30 {\AA} and with a percentage of contamination going from 55-70\% to 35\%, depending
on the mass range. The red UVJ and the red-sequence samples show a similar and smaller 
contamination, both in percentage ($\sim40$\%) and median value ($\sim$15-10 {\AA}). The less 
contaminated samples are the quiescent and the red SED galaxies: the percentage
of galaxies with emission lines $>$5 {\AA} is $\sim$30\% at low redshift and slightly smaller ($\sim$20\%)
at higher redshift, with median values closer to the adopted cut ($\sim$8-10 {\AA}).

Many works have already shown the presence of emission lines in the spectra of red/passive galaxies. However,
given their line ratios, these are usually classified as LINERS \citep[low-ionization 
nuclear emission-line regions,][]{Heckman1980}, or with a LINER-like emission 
\citep[e.g., see][]{Phillips1986,Yan2006,Annibali2010,Yan2012}. As summarized by
\cite{Annibali2010}, the most probable mechanisms proposed to produce such lines are 
a low accretion-rate AGN, fast shocks, or photoionization by old post-asymptotic giant branch 
(PAGB) stars. More recent studies favor the last option \citep[see also][]{Yan2012}.
It is beyond the aim of this paper to analyze in detail the source of ionization in this population of galaxies; what
we wanted to stress here is that in this population of galaxies weak emission lines do not
necessarily trace star-formation activity, especially if the sSFR as found from the SED-fitting
does not confirm a significant star formation. On the other hand, the stronger emission lines found in morphological ETGs
should actually trace star-formation activity, as confirmed by the ratio with the other significant
emission lines (H$\beta$, [NII]$\lambda$6583, [OII]$\lambda$5007).

\paragraph{\bf Nonelliptical morphology.} 
The median morphology estimated for the galaxies not matching an elliptical template is equal to a 
ZEST type = 2.1 in all samples, which corresponds to a bulge-dominated spiral galaxy. This means that 
for all samples the tail of morphologies is not extended toward very late-type templates.
However, all the samples show a high percentage of contamination due to non elliptical-galaxies,
with percentages going from $\sim$60\% at low masses to $\sim$25-30\% at high masses, the less
biased sample being the red \& passive ETGs one.

This evidence confirms that the color transformation and the morphological transformation
from late-type to early-type galaxies are not concomitant processes and that the quenching of the 
star-formation activity could precede the change in morphological type.

\subsection{Contamination as a function of mass}
\label{sec:resmass}
The analysis of the different properties of passive galaxies indicated a clearly evident dependence of
the contamination on the stellar mass. Looking at Tab. \ref{tab:tab2}, it is possible to see that the 
percentage of contamination in almost all the analyzed properties decreases from 
${\rm log({\mathcal M}/{\mathcal M}_{\odot})<10.25}$ to ${\rm log({\mathcal M}/{\mathcal M}_{\odot})>10.75}$ 
by a factor $\sim$2-3. Figures \ref{fig:contamination1}, \ref{fig:contamination2}, and \ref{fig:contamination3}
also highlight a mass dependence of the absolute value of the properties of the contaminants in all samples:
for example the colors of contaminants are redder by $\sim$10-30\% at high masses than at low masses and are closer to
the red sequence. This trend can also be found in sSFR and equivalent widths of emission lines,
with the lower mass bin having a star-formation activity $\sim$10\% higher and emission lines $\sim$5 {\AA}
larger than the high-mass bin for all the samples considered, both at high and at low redshift.

From this analysis we found that, irrespective of the adopted selection criterion,
an additional stellar mass cut provides a purer sample of passively evolving galaxies, 
significantly reducing the contamination by star-forming galaxies.

\begin{figure*}[t!]
\begin{center}
\includegraphics[angle=-90, width=0.85\textwidth]{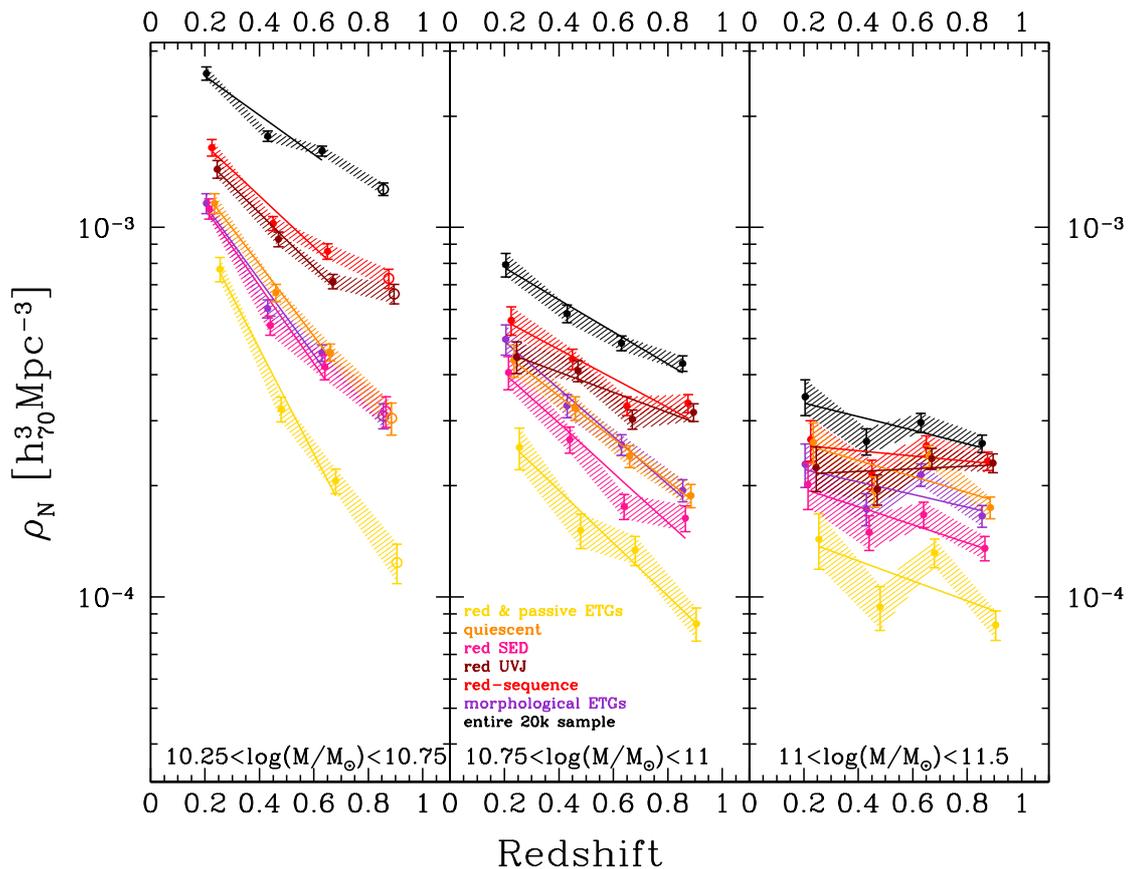}
\end{center}
\caption{Redshift evolution of the galaxy number density, derived using ${\rm1/V_{max}}$, for three different mass
ranges: ${\rm 10.25<log({\mathcal M}/{\mathcal M}_{\odot})<10.75}$ (left panel), ${\rm 10.75<log({\mathcal M}/{\mathcal M}_{\odot})<11}$ 
(central panel) and ${\rm log({\mathcal M}/{\mathcal M}_{\odot})>11}$ (right panel). The different colors represent 
different selection criteria (violet for the morphological ETGs, light red for the red-sequence galaxies, dark red for the red UVJ galaxies, 
pink for the red SED galaxies, orange for the quiescent galaxies, and yellow for the red \& passive ETGs). 
The lines represent the fit to the observed ${\rm log(\rho_{N}(z))}$ relation. The open points represent the lower limits where the survey is not 
complete.
\label{fig:numberdensity}}
\end{figure*}

\subsection{Contamination as a function of selection criteria}
\label{sec:ressel}
There are different ways to quantify the best method to select a population of passively evolving galaxies.
On one hand, it is possible to ask which criterion is least biased by the presence of star-forming
outliers, independently of how much information is used. On the other hand, at a fixed percentage 
of contamination, it is possible to ask which is the most economic criterion in terms of information
used. The first will privilege the purity of the sample but has the drawback of requiring a large amount
of data to be used. In contrast, the second will be necessarily more contaminated but also of greatest
interest in many surveys where less data are available, due to low spectral resolution or limited wavelength coverage.

Needless to say, the purest sample is the one based on the combined selection criterion since the use
of photometric, spectroscopic, and morphological information helps to minimize the presence of blue, star-forming
galaxies. It provides red and passive galaxies with a color contamination, also in the IRAC colors, $<$10\% 
for stellar masses ${\rm log({\mathcal M}/{\mathcal M}_{\odot})>10.25}$, both at low and high redshifts. The 
specific star formation of the outliers ($\lesssim$10\% for ${\rm log({\mathcal M}/{\mathcal M}_{\odot})>10.25}$)
is only $\sim$20\% higher than the chosen passive cut, and the contaminants equivalent widths of emission 
lines, both [OII] and H$\alpha$, are smaller than $\sim$10 {\AA}.

Apart from this criterion, all the others, except the morphological criterion, are equivalent in terms of requirements 
since they all need a best fit to the observed SED to obtain stellar masses, SFRs, and/or restframe colors necessary to perform the selection. 
Among them, the best-performing criteria have proven to be the red SED and the quiescent ones. An analysis of Tab. \ref{tab:tab2} and Figs. 
\ref{fig:contamination1}, \ref{fig:contamination2}, and \ref{fig:contamination3} shows that they have the minimum 
percentage of contamination in both optical colors ($\lesssim$5\%), IRAC colors ($\lesssim$15\%), spectroscopic features ($\lesssim30\%$), and sSFR
($\lesssim$15\%). The percentage of contamination is slightly 
higher than the red \& passive ETGs criterion, on average by a factor of $\sim$2, but the absolute values of the properties 
of contaminants are still compatible with a red, passively evolving population.

\subsection{Number density}
For all the samples, we estimated the galaxy stellar mass functions using the nonparametric ${\rm 1/V_{max}}$ 
formalism \citep{Schmidt1968}, from which we derived the number density ($\rho_{N}$) in three mass bins, 
${\rm log({\mathcal M}/{\mathcal M}_{\odot})=10.25-10.75, 10.75-11, 11-11.5}$. The redshift evolution of the galaxy number density is shown 
in Fig. \ref{fig:numberdensity}. As a comparison, we also reported the number density in the same mass range for the 
parent zCOSMOS-20k sample. Each relation was fitted using a weighted linear least square minimization,
considering only the redshift bins in which the data are complete. The trends are similar for the differently 
selected samples, even if the normalization is different.

In the lower stellar mass bin, we find a pronounced evolution, with an increase in number densities by a factor $\sim2-4$ between $z\sim0.65$
and $z\sim0.2$. At stellar masses ${\rm 10.75<log({\mathcal M}/{\mathcal M}_{\odot})<11}$, we still find a noticeable evolution, with an
increase by a factor $\sim$2-3 between $z\sim0.85$ and $z\sim0.2$. At higher masses, ${\rm 11<log({\mathcal M}/{\mathcal M}_{\odot})<11.5}$,
the evolution is much less pronounced, with a percentage increase in number density of $\sim$10-50\% in the entire 
redshift range.

The stronger increase in the number density of low/intermediate-mass passive galaxies with respect to massive and passive galaxies is a clear indication
of mass-assembly ``downsizing'' \citep{Fontana2004,Drory2005,Bundy2006,Cimatti2006,Thomas2010,Pozzetti2010,Moresco2010}, 
with more massive galaxies having assembled their mass at higher redshifts
and already being in place at $z\sim1$. The result of this analysis is in agreement with many other works: \cite{Scarlata2007b}
studying ETGs in the COSMOS field found no traces of significant evolution in the number density of bright ($\sim L>2.5L^{*}$) ETGs, 
with a maximum increase of $\sim$30\% from $z\sim0.7$ to $z\sim0$ when allowing for different SFHs and cosmic variance; 
\cite{Pozzetti2010} found an almost negligible evolution in the number density of quiescent galaxies in zCOSMOS-10k 
sample, $<0.1$ dex between $z=0.85$ and $z=0.25$ for ${\rm log({\mathcal M}/{\mathcal M}_{\odot})>11-11.5}$; from the analysis of the UltraVISTA-DR1 
sample \cite{Ilbert2013} found that massive galaxies (${\rm log({\mathcal M}/{\mathcal M}_{\odot})>11.2}$)
do not show any significant evolution between $0.8<z<1.1$ and $0.2<z<0.5$, while low-mass ones (${\rm log({\mathcal M}/{\mathcal M}_{\odot})\sim9.5}$)
increase in number density by a factor $>$5. \cite{Brammer2011} analyzed the number density of quiescent galaxies at $0.4<z<2.2$ 
using the NEWFIRM Medium-Band Survey, and found a strong evolution of ${\rm log({\mathcal M}/{\mathcal M}_{\odot})>10.5}$ galaxies,
a factor $\sim10$ from $z\sim2$ to $z\sim0$. However, this evolution becomes smaller and comparable with the errorbars at high masses 
between $z\sim1$ and $z\sim0.6$, only $\sim$20\% for ${\rm log({\mathcal M}/{\mathcal M}_{\odot})>11}$ ($\sim$10\% for 
${\rm 10.6<log({\mathcal M}/{\mathcal M}_{\odot})<11}$), therefore it is compatible with our results. \cite{Maraston2012},
studying the stellar mass function of BOSS galaxies, find that the galaxy number density above $\sim2.5$ $10^{11} M_{\odot}$ agrees with previous 
measurements within 2$\sigma$, with no evolution detected from $z\sim0.6$. Also from the analysis of the VIPERS survey, \cite{Davidzon2013} 
find a much stronger evolution in the number density of less massive galaxies when compared to less massive ones, with an increase of $\sim$80\%
between $z=1$ and $z=0.6$ for masses ${\rm 11<log({\mathcal M}/{\mathcal M}_{\odot})<11.4}$ and only of $\sim$45\%
between $z=1.2$ and $z=0.6$ for masses ${\rm log({\mathcal M}/{\mathcal M}_{\odot})>11.4}$. Finally, \cite{Moustakas2013} 
measured the evolution of the stellar mass function from the PRism MUlti-object Survey and confirmed an evolution by a factor of 
$3.2\pm0.5$ from $z=0.4$ to $z=0$ for stellar masses ${\rm 9.5<log({\mathcal M}/{\mathcal M}_{\odot})<10}$, by a factor of 
$2.2\pm0.4$ from $z=0.6$ to $z=0$ for stellar masses ${\rm 10<log({\mathcal M}/{\mathcal M}_{\odot})<10.5}$, and only 
$\sim58\%\pm9\%$ for stellar masses ${\rm 10.5<log({\mathcal M}/{\mathcal M}_{\odot})<11}$.

The comparison, at fixed mass, of the number density of the differently selected passive galaxies is also interesting since it sheds some
light on the timescales characteristic of different processes. At masses ${\rm log({\mathcal M}/{\mathcal M}_{\odot})<11}$,
we find that the number densities of the various samples are well separated, with highest normalization for color-selected ETGs,
intermediate for the passive/red-SED/morphologically selected ETGs, and lowest for the red \& passive ETGs. This may suggest a scenario
during which, at these masses, the first transition that an ETG experiences is a color transformation from blue to red, followed by the quenching of
the star formation and by the morphological transformation into ellipticals. 

A similar trend is found in the work of \cite{Pozzetti2010}, which quantified the timescale of the delay between color and morphological 
transformation to be around 1-2 Gyr. Some support to this scenario also comes from the work of \cite{Ilbert2010}, who find that the stellar 
mass density of quiescent galaxies is always higher than the one of red elliptical galaxies. Obviously the quantitative estimate
of the delay between these processes also depends on the adopted cuts to select ETGs since, for example, a lower/higher cut in sSFR will
shift the relation lower/higher.

\subsection{Revising the color selection criteria}
\label{sec:newcols}
In Sect. \ref{sec:col} and \ref{sec:contamination} we presented and discussed the fact that the red-sequence and the red UVJ
samples have a higher contamination, which is also clearly evident by a visual inspection of the color-mass and color-color diagrams.
However, we also argued that this higher contamination may be due to the fact that the proposed cut may not be optimal, at least for the 
zCOSMOS-20k survey.

\begin{figure}[t!]
\begin{center}
\includegraphics[angle=-90, width=0.45\textwidth]{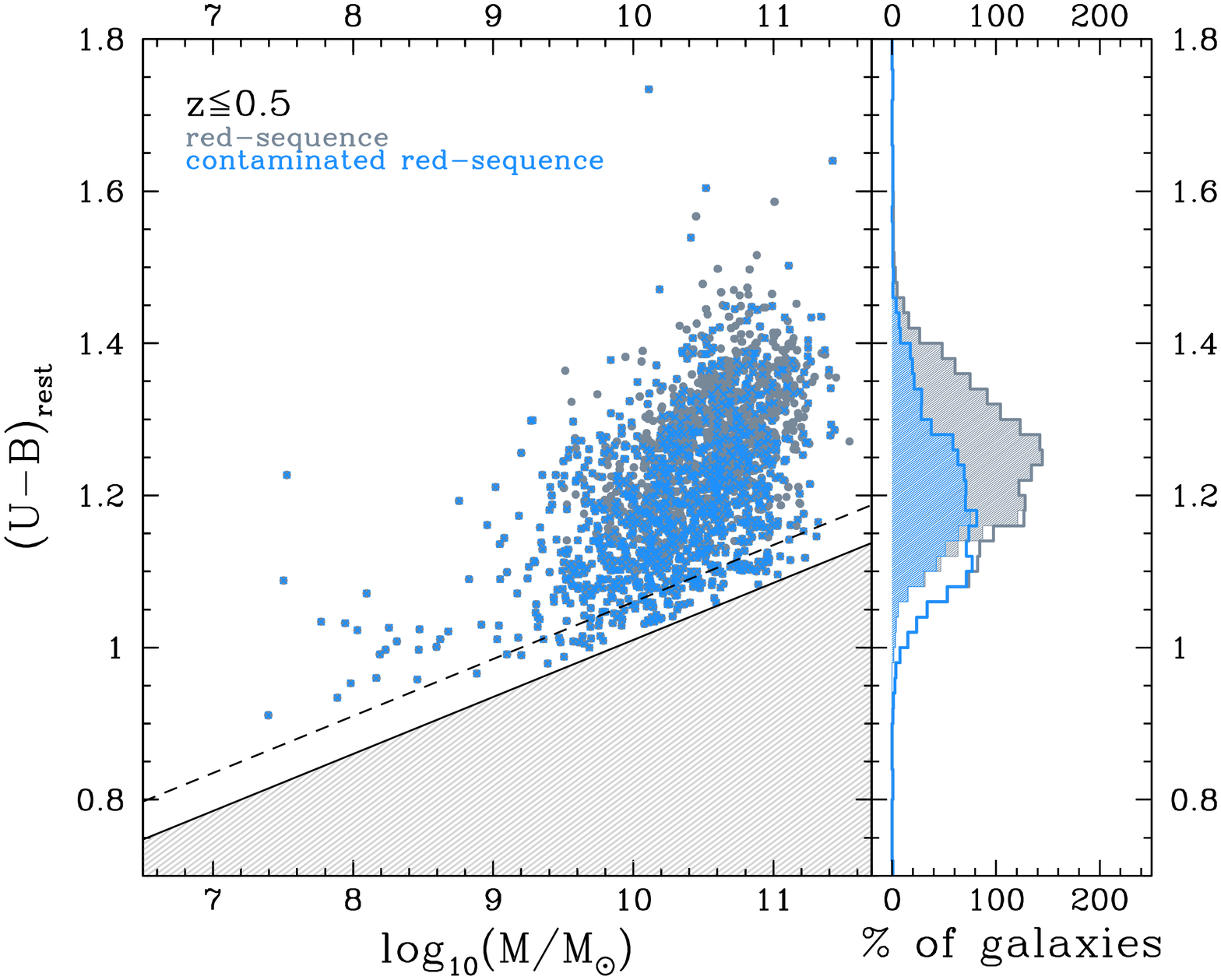}
\includegraphics[angle=-90, width=0.45\textwidth]{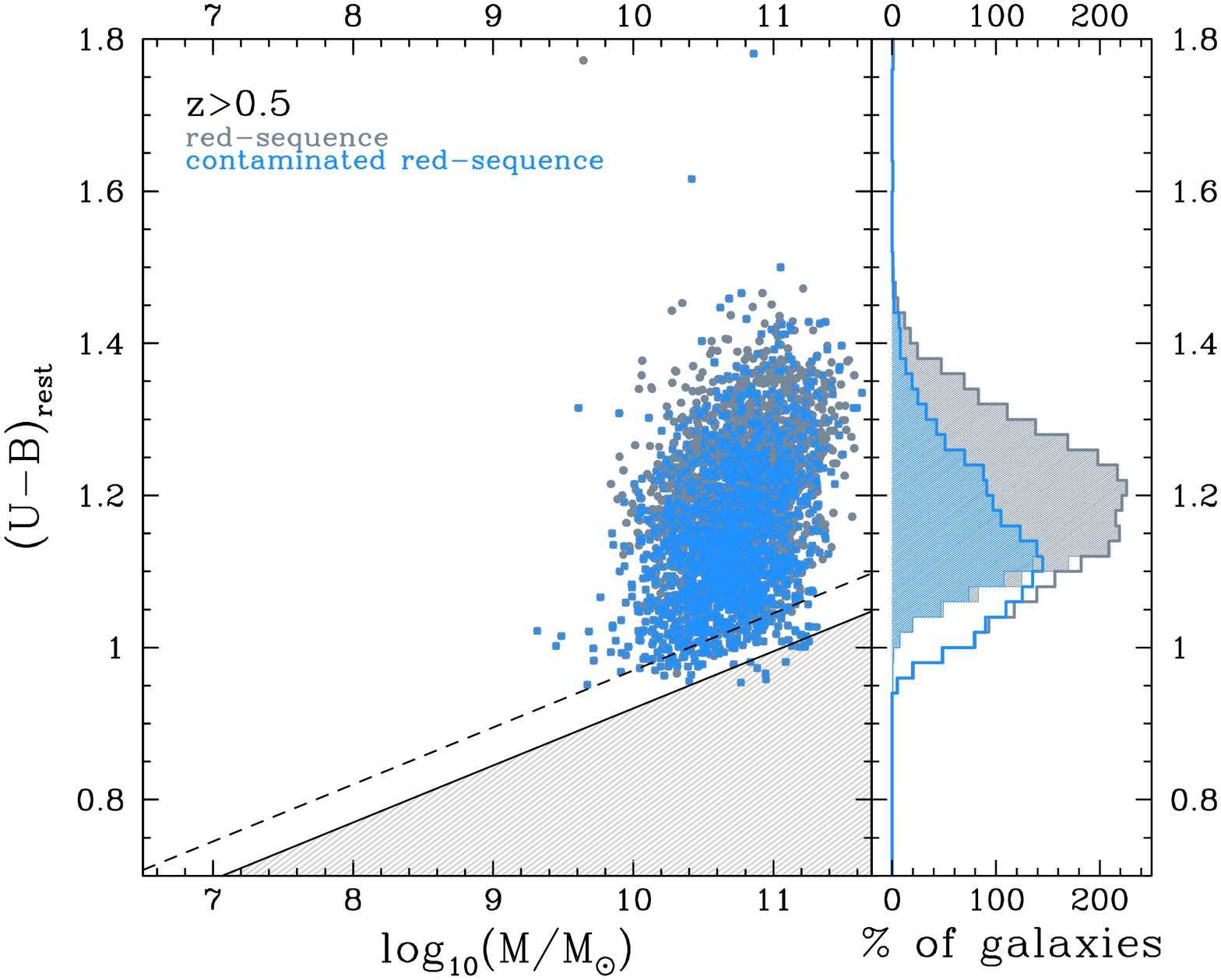}
\end{center}
\caption{$(U-B)_{\rm rest}$-mass diagram. In gray is shown the original $(U-B)_{\rm rest}$-mass criterion 
as defined by \cite{Peng2010}, and in cyan are highlighted the galaxies presenting a sSFR or emission 
lines in excess with respect to the passive cuts. The empty cyan histograms show the distribution of the 
contaminants with the old definitions, while the shaded cyan histograms show the distribution of the contaminants 
with the new selection criterion as defined in section \ref{sec:newcols}. The upper plot shows the diagram 
obtained for $z\leq0.5$, and the lower plot shows the diagram obtained for $z>0.5$.
\label{fig:newcolors1}}
\end{figure}

\begin{figure}[t!]
\begin{center}
\includegraphics[angle=-90, width=0.45\textwidth]{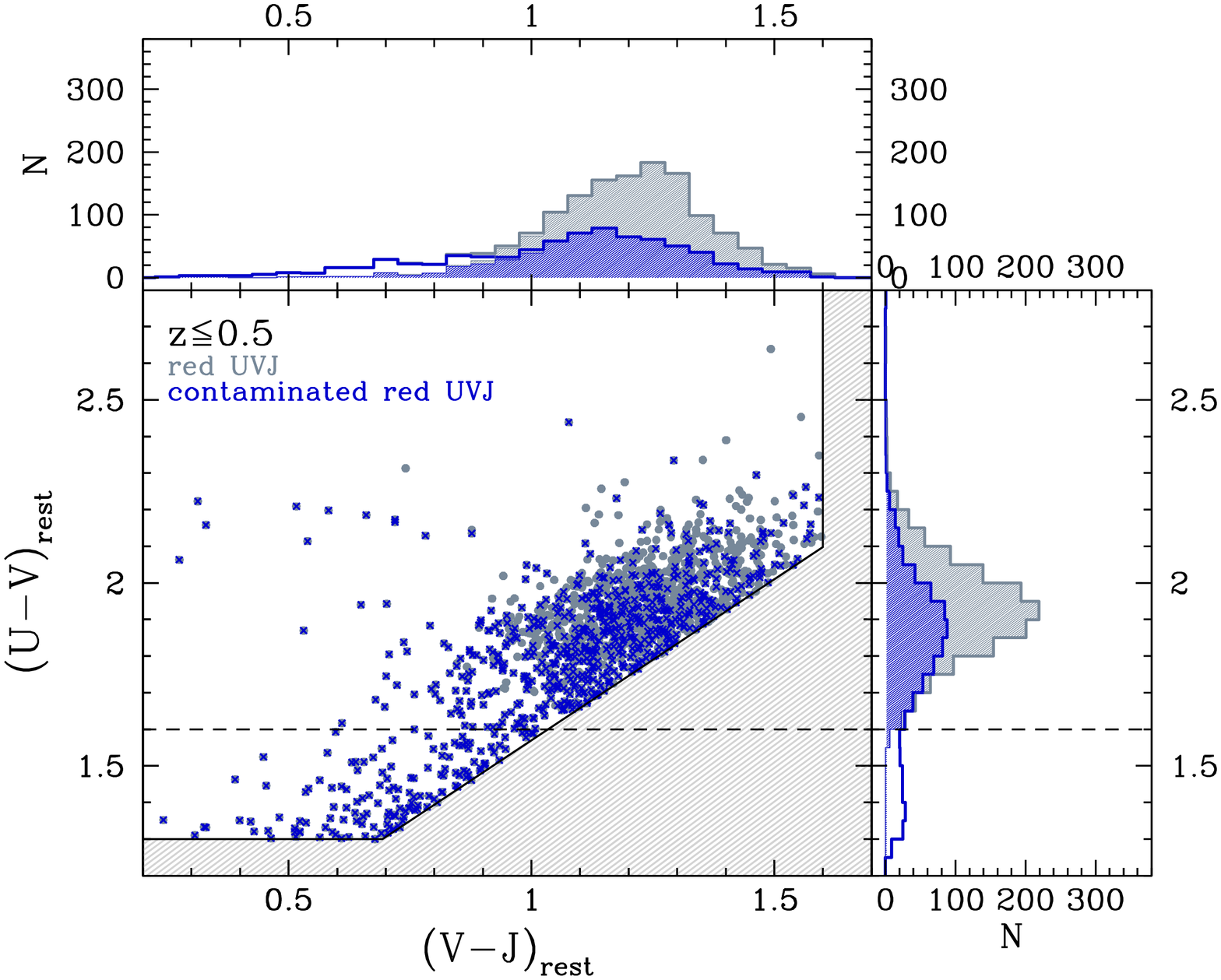}
\includegraphics[angle=-90, width=0.45\textwidth]{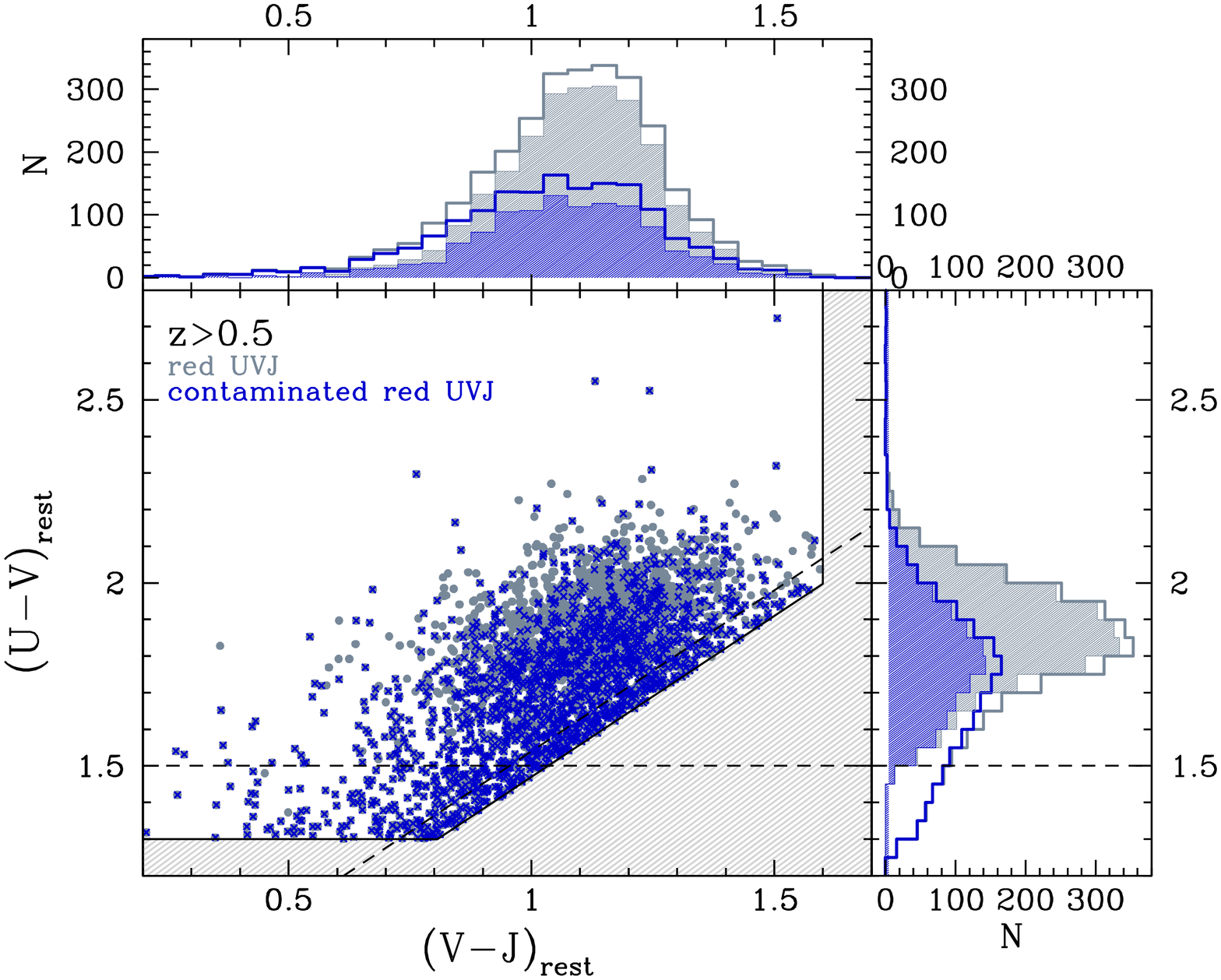}
\end{center}
\caption{$(U-V)_{\rm rest}$-$(V-J)_{\rm rest}$ diagram. In gray is shown the original UVJ criterion,
as defined by \cite{Williams2009}, and in blue are highlighted the galaxies presenting a sSFR or emission 
lines in excess with respect to the passive cuts. The empty blue histograms show the distribution of the 
contaminants with the old definitions, while the shaded blue histograms show the distribution of the contaminants 
with the new selection criterion as defined in section \ref{sec:newcols}. The upper plot shows the diagram obtained 
for $z\leq0.5$, and the lower plot shows the diagram obtained for $z>0.5$.
\label{fig:newcolors2}}
\end{figure}

To address this issue, we decided to study how the colors of the contaminants (as defined in Sect. \ref{sec:contamination}) for these
two criteria are distributed in the $(U-B)_{\rm rest}$ color-mass and UVJ color-color diagrams, respectively, in order to understand if an additional cut is able
to further reduce the contamination. In Fig. \ref{fig:newcolors1} we show the $(U-B)_{\rm rest}$-mass diagram for the red-sequence
galaxies (in gray), highlighting in cyan the galaxies presenting a sSFR or emission lines in excess with respect to the passive cuts defined (i.e., the {\it contaminants});
Fig \ref{fig:newcolors2} shows the same as Fig. \ref{fig:newcolors1}, but for the $(U-V)_{\rm rest}$-$(V-J)_{\rm rest}$ diagram and for the case of the 
red UVJ galaxies. From these plots it is evident that statistically the contaminants have bluer colors and therefore that a refined cut may help
to reduce the contamination. In particular from Fig. \ref{fig:newcolors2} it is clear that there are specific ranges of colors ($1.3<(U-V)_{\rm rest}<1.5$)
where the contaminants represent 100\% of the distribution.

We therefore defined two revised color selection criteria, with slightly redder colors: for the $(U-B)_{\rm rest}$-mass diagram:
\begin{equation}
(U-B)_{\rm rest}>1.15+0.075\times log(\frac{\mathcal M}{10^{10}\mathcal M_{\odot}})-0.18\times z\nonumber
\end{equation}
and for the $(U-V)_{\rm rest}$-$(V-J)_{\rm rest}$ diagram:
\begin{equation}
  \left\{
    \begin{array}{ll}
      (U-V)_{\rm rest}>1.6 &\\
      (V-J)_{\rm rest}>1.6 & [0<z<0.5]\\
      (U-V)_{\rm rest}>0.88*(V-J)_{\rm rest}+0.69\nonumber
    \end{array}
    \right.
\end{equation}
\begin{equation}
  \left\{
    \begin{array}{ll}
      (U-V)_{\rm rest}>1.5 &\\
      (V-J)_{\rm rest}>1.6 & [0.5<z<1]\\
      (U-V)_{\rm rest}>0.88*(V-J)_{\rm rest}+0.66\nonumber
    \end{array}
    \right.
\end{equation}
The empty colored histograms in Figs. \ref{fig:newcolors1} and \ref{fig:newcolors2} show the distribution of the contaminants with the old definitions,
while the shaded colored histograms show the distribution of the contaminants obtained with the new definitions just reported. It is evident how these 
new color selection criteria help to cut significantly the contaminants with blue colors.

We checked that these new selection criteria provide purer samples than the previous ones. The new red UVJ sample no longer presents significant blue
tails in the color-mass diagram, having a contamination compatible with all the other samples. In addition, the percentage of contamination of
the new red-sequence sample in the UVJ color-color diagram is reduced from 20-40\% to 15-25\% at low redshift and from 20-40\% to 10-15\% at high redshift. 
However, even if reduced with these new definitions, the contamination in terms of sSFR and presence of emission lines is still higher than the quiescent, 
red SED, and red \& passive ETGs criteria, which continue to behave better than the color criteria in selecting a purer sample 
of ETGs.

\section{Summary and conclusions}\label{sec:concl}

We have studied the photometric, spectroscopic and morphological properties of six differently selected samples of passive galaxies 
in order to analyze their contamination in terms of blue/star-forming/nonpassive galaxies. From the zCOSMOS-20k spectroscopic
catalog, we extracted a sample based on morphology (3336 ``morphological'' early-type galaxies), on optical colors (4889 
``red-sequence'' and 4882 ``red UVJ'' galaxies), on specific star-formation rate (2937 ``quiescent'' galaxies), 
on a best fit to the observed SED (2603 ``red SED'' galaxies), and on a criterion that combines morphological, 
spectroscopic, and photometric information (1530 ``red \& passive early-type galaxies''). 

For all the samples, we estimated the $(U-B)_{\rm rest}$-stellar mass 
diagram, the IRAC color-color plot as defined by \cite{Lacy2004}, the sSFR-${\rm EW_{0}([OII]/H\alpha)}$ diagram, and the 
morphological types. We also evaluated the median stacked spectra of these samples to search in detail for the presence 
of peculiar spectroscopic features that may be a proxy of star-formation activity.

To quantify the contamination of the various samples,  for each property we defined a passive cut and studied 
the percentage of galaxies not fulfilling this cut, as well as the median value of the parameters (color, sSFR, EW$_{0}$ of
emission lines) involved in each selection. In this way we were also able to study
how much these contaminants are close to or far from the assumed passive definition. All these analyses were
performed in two redshift bins ($z\leq0.5$ and $z>0.5$) and three stellar mass bins (${\rm log({\mathcal M}/{\mathcal M}_{\odot})<10.25}$, 
${\rm 10.25<log({\mathcal M}/{\mathcal M}_{\odot})<10.75}$, and ${\rm log({\mathcal M}/{\mathcal M}_{\odot})>10.75}$).

Our main results are listed below:
\begin{itemize}
\item We find that all samples display tails in the star-forming regions of different diagrams, irrespective of the severity of the
criterion adopted. This contamination has been shown to be dependent on the stellar mass for all the samples, with more 
massive samples being less biased than less massive ones. This fact demonstrates that a cut in stellar mass is helpful to improve
the purity of the sample.
\item The comparison between the different selection criteria shows that the best performing one is based on a combined selection
since it takes into account all the available information about the galaxies (morphological, spectroscopic and photometric), with
the obvious disadvantage of being highly demanding in terms of the data needed. For this sample, we find that the contamination
is minimized, especially for stellar masses ${\rm log({\mathcal M}/{\mathcal M}_{\odot})>10.25}$: $<$10\% for both optical and IRAC 
colors, $\lesssim$10\% in sSFR and with weak emission line equivalent widths for the contaminants ($\lesssim$10 {\AA}).
\item The morphological criterion is the one with the larger contamination, and a high percentage of blue, emission-line, star-forming
elliptical galaxies has been found in this sample ($\sim$12-65\%, depending on the mass range). Among the other selection
criteria, we identify the red SED and the quiescent as the best since their slight increase in 
contamination (about a factor 2) is offset by the fact that they are extremely economical in terms of information required. 
\item The slope of the number density as a function of redshift at fixed mass for the various samples is similar. We found an increase 
in number density in the mass range 
${\rm 10.25<log({\mathcal M}/{\mathcal M}_{\odot})<10.75}$ from $z=0.65$ to $z=0.2$ by a factor $\sim$2-4, in the mass range 
${\rm 10.75<log({\mathcal M}/{\mathcal M}_{\odot})<11}$ from $z=1$ to $z=0.2$ by a factor $\sim$2-3, while for massive ETGs
${\rm 11<log({\mathcal M}/{\mathcal M}_{\odot})<11.5}$ this increase from $z=1$ to $z=0.2$ is only $\sim$10-50\% (and $<$10\% between
$z=1$ and $z=0.4$). This trend in mass confirms that most massive ETGs are already in place at $z\sim1$. The comparison
of the different number densities suggests a scenario for ETGs in which the color transformation from blue to red precedes the quenching of the star 
formation and the morphological transformation from late type to ellipticals.
\item By analyzing the color-mass and color-color diagrams, we proposed two revised selection criteria, which help to reduce
the contamination by blue star-forming galaxies.
\end{itemize}

Selecting a sample of passive galaxies that is as uncontaminated as possible by star-forming objects is crucial for an unbiased study of the properties 
of this population of galaxies. The analysis presented in this paper provides a first detailed comparison of the contamination obtained 
with different selection criteria, and represents an important step forward in the identification of the best ways of selecting passive galaxies. 
This will be especially fundamental for the new and upcoming surveys such as BOSS \citep{Schlegel2009} and Euclid \citep{Laureijs2011}, 
which will provide the scientific community unprecedented statistics of these objects, in particular at high redshift. From this
point of view, this work represents the starting point to study and develop optimized passive galaxy selection criteria also at $z>1$, which could be applied,
for example, to Euclid simulations.

In a forthcoming paper
we aim to study how much the different selections presented here affect the study of the evolution of passive galaxies.

\begin{acknowledgements}
MM and AC acknowledge the grants ASI n.I/023/12/0 ``Attivit\'a relative alla fase B2/C
per la missione Euclid'' and MIUR PRIN 2010-2011 ``The dark Universe and the 
cosmic evolution of baryons: from current surveys to Euclid''. Part of the work has
also been supported by an INAF grant ``PRIN-2010''. This analysis is based on zCOSMOS observations, 
carried out using the Very Large Telescope at the ESO Paranal Observatory under Programme ID: LP175.A-0839.
We would like to thank the anonymous referee for the useful comments and suggestions, which helped to improve the paper.
\end{acknowledgements}

\clearpage

\end{document}